\documentclass[fleqn,usenatbib]{mnras}

\usepackage[T1]{fontenc}

\DeclareRobustCommand{\VAN}[3]{#2}
\let\VANthebibliography\thebibliography
\def\thebibliography{\DeclareRobustCommand{\VAN}[3]{##3}\VANthebibliography}

\usepackage{graphicx}	
\usepackage{amsmath}	
\usepackage{amssymb}	
\usepackage{newtxtext,newtxmath}

\usepackage[usenames]{color}
\usepackage{natbib}

\newcommand{\Msol}{$M_{\sun}$}
\newcommand{\arcdeg}{$^{\circ}$}

\newcommand{\kms}{\,km\,s$^{-1}$}

\title[A GMC catalog of Centaurus\,A]{A Giant Molecular Cloud Catalog in the Molecular Disk of the Elliptical Galaxy NGC\,5128 (Centaurus\,A)}

\author[R. E. Miura et al$^{1}$]{
R. E. Miura,$^{1}$\thanks{E-mail: rie.miura@nao.ac.jp}
D. Espada,$^{2,3}$
A. Hirota,$^{1,4}$
C. Henkel,$^{5,6}$
S. Verley,$^{3,7}$
M. I. N. Kobayashi,$^{8,9}$
S. Matsushita,$^{10}$
\newauthor
F. P. Israel,$^{11}$
B. Vila-Vilaro,$^{4,12}$
K. Morokuma-Matsui,$^{13,14}$
J. Ott,$^{15}$
C. Vlahakis,$^{16}$
A. B. Peck,$^{17}$
S. Aalto,$^{18}$
\newauthor
M. R. Hogerheijde,$^{19,20}$
N. Neumayer,$^{21}$
D. Iono,$^{1,22}$
K. Kohno,$^{14,23}$
H. Takemura,$^{1,22}$
S. Komugi$^{24}$
\\
$^{1}$National Astronomical Observatory of Japan,
National Institutes of Natural Sciences,
2-21-1 Osawa, Mitaka, Tokyo 181-8588, Japan\\
$^{2}$SKA Organisation, Jodrell Bank, Lower Withington, Macclesfield, Cheshire SK11 9FT, UK\\
$^{3}$Departamento de Física Teórica y del Cosmos, Campus de Fuentenueva, Universidad de Granada, E18071–Granada, Spain\\
$^{4}$Joint ALMA Observatory, Alonso de C{\'o}rdova, 3107, Vitacura, Santiago 763-0355, Chile\\
$^{5}$Max-Planck-Institut f{\"u}r Radioastronomie, Auf dem H{\"u}gel 69, 53121, Bonn, Germany\\
$^{6}$Dept. of Astronomy, King Abdulaziz University, PO Box 80203, 21589 Jeddah, Saudi Arabia\\
$^{7}$Instituto Universitario Carlos I de F{\'i}sica Te{\'o}rica y Computacional, Facultad de Ciencias, E-18071 Granada, Spain\\
$^{8}$Department of Earth and Space Science, Graduate School of Science, Osaka University, 1-1 Machikaneyama-cho, Toyonaka, Osaka 560-0043, Japan\\
$^{9}$Astronomical Institute, Tohoku University, Aoba, Sendai, Miyagi, 980-8578, Japan\\
$^{10}$Academia Sinica Institute of Astronomy and Astrophysics, 11F of Astro-Math Bldg, AS/NTU, No.1, Section 4, Roosevelt Rd, Taipei 10617, Taiwan, Republic of China\\
$^{11}$Sterrewacht Leiden, Leiden University, PO Box 9513, 2300 RA, Leiden, The Netherlands\\
$^{12}$European Southern Observatory, Alonso de C{\'o}rdova 3107, Vitacura, Santiago, Chile\\
$^{13}$Institute of Space and Astronautical Science, Japan Aerospace Exploration Agency, 3-1-1 Yoshinodai, Chuo-ku, Sagamihara, Kanagawa 252-5210, Japan\\
$^{14}$Institute of Astronomy, School of Science, The University of Tokyo, 2-21-1 Osawa, Mitaka, Tokyo 181-0015, Japan\\
$^{15}$National Radio Astronomy Observatory, PO Box O, 1003 Lopezville Road, Socorro, NM 87801, USA\\
$^{16}$National Radio Astronomy Observatory, 520 Edgemont Road, Charlottesville, VA 22903-2475, USA\\
$^{17}$Gemini Observatory, 670 N'Aohoku Pl, Hilo 96720-2700, Hawaii, HI, USA\\
$^{18}$Dep. of Space, Earth and Environment, Chalmers University of Technology, Onsala Space Observatory, SE-43992 Onsala, Sweden\\
$^{19}$Leiden Observatory, Leiden University, PO Box 9513, 2300 RA Leiden, The Netherlands\\
$^{20}$Anton Pannekoek Institute for Astronomy, University of Amsterdam, Science Park 904, 1098 XH, Amsterdam, The Netherlands\\
$^{21}$Max Planck Institute for Astronomy (MPIA), K{\"o}nigstuhl 17, D-69121 Heidelberg, Germany\\
$^{22}$The Graduate University for Advanced Studies (SOKENDAI), 2-21-1 Osawa, Mitaka, Tokyo, 181-0015, Japan\\
$^{23}$Research Center for the Early Universe, School of Science, The University of Tokyo, 7-3-1 Hongo, Bunkyo-ku, Tokyo 113-0033, Japan\\
$^{24}$Division of Liberal Arts, Kogakuin University, 2665-1, Hachioji, Tokyo 192-0015, Japan
}

\date{Accepted XXX. Received YYY; in original form ZZZ}

\pubyear{2020}

\begin{document}
\label{firstpage}
\pagerange{\pageref{firstpage}--\pageref{lastpage}}
\maketitle

\begin{abstract}
  We present the first census of giant molecular clouds (GMCs) complete down to 10$^6$\,\Msol\ and within the inner 4\,kpc of the nearest giant elliptical and powerful radio galaxy, Centaurus~A. 
  {  We identified 689 GMCs using CO(1--0) data with 1\arcsec\ spatial resolution ($\sim 20$\,pc) and 2\,\kms\ velocity resolution obtained with the Atacama Large Millimeter/submillimeter Array (ALMA).
   The $I$(CO)-$N$(H$_2$) conversion factor based on the virial method is $X_{\rm CO}$ = $(2 \pm 1 )\times10^{20}$\,cm$^{-2}$(K\,\kms)$^{-1}$ for the entire molecular disk, consistent with that of the disks of spiral galaxies including the Milky Way, and $X_{\rm CO}$ = $(5 \pm 2)\times10^{20}$\,cm$^{-2}$(K\,\kms)$^{-1}$ for the circumnuclear disk (CND, within a galactocentric radius of 200\,pc).
     We obtained the GMC mass spectrum distribution and find that the best truncated power-law fit for the whole molecular disk, with index $\gamma \simeq -2.41 \pm 0.02$ and upper cutoff mass $\sim1.3\times10^{7}\,M_{\sun}$, is also in agreement with that of nearby disk galaxies.  A trend is found in the mass spectrum index from steep to shallow as we move to inner radii.
  Although the GMCs are in an elliptical galaxy, the general GMC properties in the molecular disk are as in spiral galaxies.   However, in the CND, large offsets in the line-width-size scaling relations ($\sim$ 0.3\,dex higher than those in the GMCs in the molecular disk), a different $X_{\rm CO}$ factor, and the shallowest GMC mass distribution shape ($\gamma = -1.1 \pm 0.2$) all suggest that there the GMCs are most strongly affected  by the presence of the AGN and/or shear motions.}

\end{abstract}




\section{Introduction}
\label{intro}
Giant Molecular Cloud (GMC) properties and the scaling relations are in general compatible in different regions of the Milky Way disk and in other galaxies across a wide range of environments \citep[e.g.][]{2008ApJ...686..948B}.
The properties of the molecular clouds depend primarily on the balance between their kinetic and gravitational potential energy, and  in general molecular clouds are seen to be bound elements with velocity dispersions counter-balancing self-gravity, as seen in the Milky Way  \citep{2009ApJ...699.1092H},  nearby dwarf galaxies \citep{2008ApJ...686..948B}, the Large Magellanic Cloud \citep[LMC,][]{2011ApJS..197...16W}, nearby spiral galaxies \citep[e.g.][]{2007ApJ...661..830R, 2012ApJ...761...37M, 2013ApJ...772..107D,2014A&A...567A.118D,2014ApJ...784....3C,2018PASJ...70...73H,0004-637X-857-1-19}, and starbursts \citep[e.g.][]{2005ApJ...623..826R,2015ApJ...801...25L}.
{ In environments with high ambient pressure and/or strong interstellar radiation field such as in the Galactic Center and/or starburst (SB) regions \citep{2001ApJ...562..348O,2015ApJ...801...25L,2018ApJ...864..120M}, the GMCs can be characterized by velocity widths which are 0.5--1\,dex higher than the average in the disks of spiral galaxies. If the surface densities of the clouds are high enough to balance the collapse due to gravitational potential and the internal pressure, they may be found as gravitationally bound entities \citep{2015ApJ...801...25L, 2018ApJ...860..172S}. On the other hand, in the low density regime,  external pressure may be needed to play a role in confining the GMCs \citep{2001ApJ...562..348O}.}

Early-type galaxies are thus good candidates to find potential differences in the GMC properties because of the higher stellar surface densities, interstellar radiation, and more diverse multi-phase interstellar medium (ISM). 
Unfortunately, the identification and study of GMCs with high resolution in early-type galaxies is largely missing.
An exception is the high angular resolution study ($\sim$ 20\,pc) of the lenticular galaxy NGC\,4526. \citet{2015ApJ...803...16U} found that although GMCs are gravitationally bound in this object, they are denser, more luminous, and exhibit greater velocity dispersions than similarly sized Galactic GMCs.
However, additional studies to resolve GMCs in other early-type galaxies are still needed.

A natural step is to study the parameter space of GMCs within the environments of a giant elliptical galaxy. This kind of study in massive elliptical galaxies has been hampered because they are less frequently found nearby, contain significantly less molecular gas, and the distribution is more compact than similarly sized spiral galaxies. However, in some cases the molecular gas in elliptical galaxies present rotating disk-like structures along their optical major axes \citep{2002AJ....124..788Y}.
Most of the published interferometric observations of molecular gas in elliptical galaxies
\citep[e.g.][]{2011MNRAS.410.1197C,2013MNRAS.432.1796A} have not been of sufficient spatial resolution and sensitivity  to address the detailed GMC properties as performed for nearby disk galaxies.

The Atacama Large Millimeter/submillimeter Array (ALMA) is providing an excellent view on what the molecular properties of truly elliptical galaxies are \citep[e.g.][]{2017ApJ...845..170B,2018ApJ...858...17T,2019ApJ...870...39V}. Many of the studied objects are thought to have their gas re-accreted by gas rich mergers later in their evolution.
CO line widths are seen to be broader ($\gtrsim$10 times) in group-centered elliptical galaxies than Galactic molecular clouds \citep{2018ApJ...858...17T}.
Molecular gas filaments are seen in the central $\sim$6.5 kpc of the elliptical NGC\,1275, probably representing pressure-confined structures created by turbulent flows \citep{2017ApJ...850...31L}.
However, these observations have not been able to resolve (spatially and kinematically), and with sufficient signal to noise, a sufficiently large number of GMCs in elliptical objects. 

Here we present the GMC properties within the molecular disk of the closest giant elliptical galaxy, NGC\,5128, which is the host of the radio-source Centaurus~A (hereafter Cen\,A). 
Cen~A is at a distance of only D $\simeq$ 3.8\,Mpc (\citealt{2010PASA...27..457H}, 1\arcsec = 18\,pc) and it is therefore by far the most adequate target in the class of giant elliptical galaxies as well as powerful radio galaxies for studies of their molecular gas with high resolution.
Indeed, Cen~A is a peculiar case of an elliptical galaxy whose gaseous component has been supplied a few 0.1\,Gyr ago by the accretion of a HI rich galaxy { \citep[e.g.][]{2010A&A...515A..67S}}.
Along the dust lane of the elliptical galaxy there is a molecular gas component of mass $\sim$ 10$^{9}$\,$M_\odot$ as probed by various molecular lines \citep[e.g.][]{1987ApJ...322L..73P,1990ApJ...365..522E,1993A&A...270L..13R,2001A&A...371..865L,2009ApJ...695..116E,2013ASPC..476...69E,2017ApJ...851...76M}, partially seen in the form of kpc scale spiral features \citep{2012ApJ...756L..10E}. { The dust lane is along the minor axis, different to other ellipticals where disks are usually along the major axis \citep{2002AJ....124..788Y}}. The molecular gas is associated with other components of the interstellar medium, such as
 ionized gas traced by the H$\alpha$ line (e.g. \citealt{1992ApJ...387..503N}), near infra-red continuum \citep{1993ApJ...412..550Q}, submillimeter continuum \citep[e.g.][]{1993MNRAS.260..844H,2002ApJ...565..131L}, and mid-IR continuum emission \citep[e.g.][]{1999A&A...341..667M,2006ApJ...645.1092Q}. In the inner hundreds of parsecs there is a circumnuclear disk (CND) of  400\,pc total extent ($\sim 24\arcsec$) and a P.A. = $155^{\circ}$, perpendicular to the inner jet, at least as seen in projection \citep{2009ApJ...695..116E}. The total gas mass in this component has been estimated to be 9 $\times$ 10$^7\,M_{\sun}$ \citep{2014A&A...562A..96I,2017A&A...599A..53I}. 
More detailed studies of the CND with higher resolutions of $\sim$ 5\,pc in CO(3--2) and CO(6--5) have revealed the complexity of the molecular gas distribution and kinematics in that region, with multiple internal filaments and shocks \citep{2017ApJ...843..136E}.

 Due to the same origin of the gas in the extended disk and in the CND as a result of the galaxy accretion, the properties of the ISM are probably similar, and likely different from those of late-type spiral galaxies.  For example, a nearly constant metallicity is found with radius \citep{2017A&A...599A..53I}. By comparing with PDR models, it is inferred that the far UV radiation field strength varies from 55 to 550\,G$_{\rm 0}$ (a measure of the strength of the FUV radiation field normalized to the Habing field, see \citealt{1968BAN....19..421H}), and total hydrogen densities vary
 between 500 and 5000\,cm$^{-3}$. The emission line properties throughout the disk of Cen\,A are similar to those in spiral galaxies at least to a first approximation \citep{2014ApJ...787...16P}. Nevertheless, the central gas probably differs from the more extended component due to its proximity to the AGN and shear motions may be stronger there.
 An estimate for the average gas to dust mass ratio is around 100, albeit for the CND it is larger $\sim$275 \citep{2012MNRAS.422.2291P,2017A&A...599A..53I}. This is probably due to dust sputtering produced by X-rays in the central regions or dust reduction close to the jets \citep{2012MNRAS.422.2291P}.

 In this paper we aim at providing a census of the GMCs as traced by CO(1--0) { down to GMC masses of 10$^5$\,$M_\odot$ and within the inner 4\,kpc} of an elliptical galaxy, from the tenuous outskirts of its molecular disk to molecular clumps close to the powerful AGN, using high resolution ($\sim$ 20\,pc), sensitivity (10\,mJy/beam in 2\,\kms\ channels), and dynamic range observations obtained with ALMA.
 The observations were presented in \citet{2019ApJ...887...88E} (Paper\,{\sc i}) in the context of a study of the star formation (SF) law across the molecular disk of Cen\,A.
 {The outline of this paper is the following. The observations as well as the
 data reduction are presented in \S\,\ref{obs}. In \S\,\ref{methods}, we show the methods for the identification of GMCs and estimation of parameters in the CO(1--0) GMC catalog.
  In \S\,\ref{results}, { we present} the main GMC properties and derive scaling relations, { which we compare} with similar studies of other galaxies from the literature.
 We also provide a measure of the $X_{\rm CO}$ conversion factor using the virial method for the entire molecular disk and also for the CND.
 In \S\,\ref{discussion}, we discuss the large $X_{\rm CO}$ found toward the CND, study the stability and pressure balance of the GMCs, calculate their virial parameters, and obtain GMC mass spectra for different regions within the molecular disk, which we compare with other observational studies and numerical calculations.}

 \section{Observations and Data Reduction}\label{obs}

 We present observations of the CO(1--0) line ($\nu_{\rm rest}$=115.271\,GHz) for a mosaic region, 5\arcmin $\times$1\farcm4 with a P.A. (North to East) of 120\arcdeg , covering the dust lane of Cen~A. The datasets were obtained as part of program 2013.1.00803.S (P.I. D. Espada). The observing setup, datasets, and calibration strategy were already introduced in Paper\,{\sc i}, so for more information please refer to that paper. Here we only provide a summary.

 CO(1--0) line data were obtained with the 12m, 7m and  Total Power (TP) arrays and therefore the final combined maps have information from small to large spatial scales.
 The observations were carried out with a Nyquist sampled configuration of 46 pointings in the 12m array and 19 in the 7m array. The half power beam width (HPBW) at 115\,GHz is $50\farcs6$ and $86\farcs8$ for a 12m and 7m antenna, respectively. The TP raster map covered a field of $405\arcsec \times 189\arcsec$.

 The calibration of the data was performed with the Common Astronomy Software Applications package \citep[CASA;][]{2007ASPC..376..127M}.
 Each of the interferometric datasets was calibrated independently and concatenated after subtracting line-free continuum emission.
 We generated a CO data cube limiting the velocity range between 242 -- 820\,\kms\ with 2.0\,\kms\ resolution using {\sf TCLEAN} task in CASA 5.4, Briggs weighting and a robust parameter of 0.5.
 Finally, the mosaicked CO(1--0) interferometric 12m plus 7m data cube was later combined using feathering with the TP cube.
 
 The total flux of the final image is 9690\,Jy\,\kms .  The CO(1--0) total flux in a region of 116\arcsec $\times$ 45\arcsec along a P.A. of 125\arcdeg\ as probed by \citet{2014A&A...562A..96I} agrees to within 10\% ($\sim$ 4500\,Jy\,\kms ). 
 The CO(1--0) cube has a typical noise level of 10\,mJy\,beam$^{-1}$ per 2\,\kms channel.
 The angular resolution of the final images is 1\farcs36 $\times$ 1\farcs03 (or 24$\times$20\,pc), with a P.A. of 61\fdg3 (HPBW).

\begin{figure*}
 \begin{center}
 \includegraphics[width=0.9\textwidth]{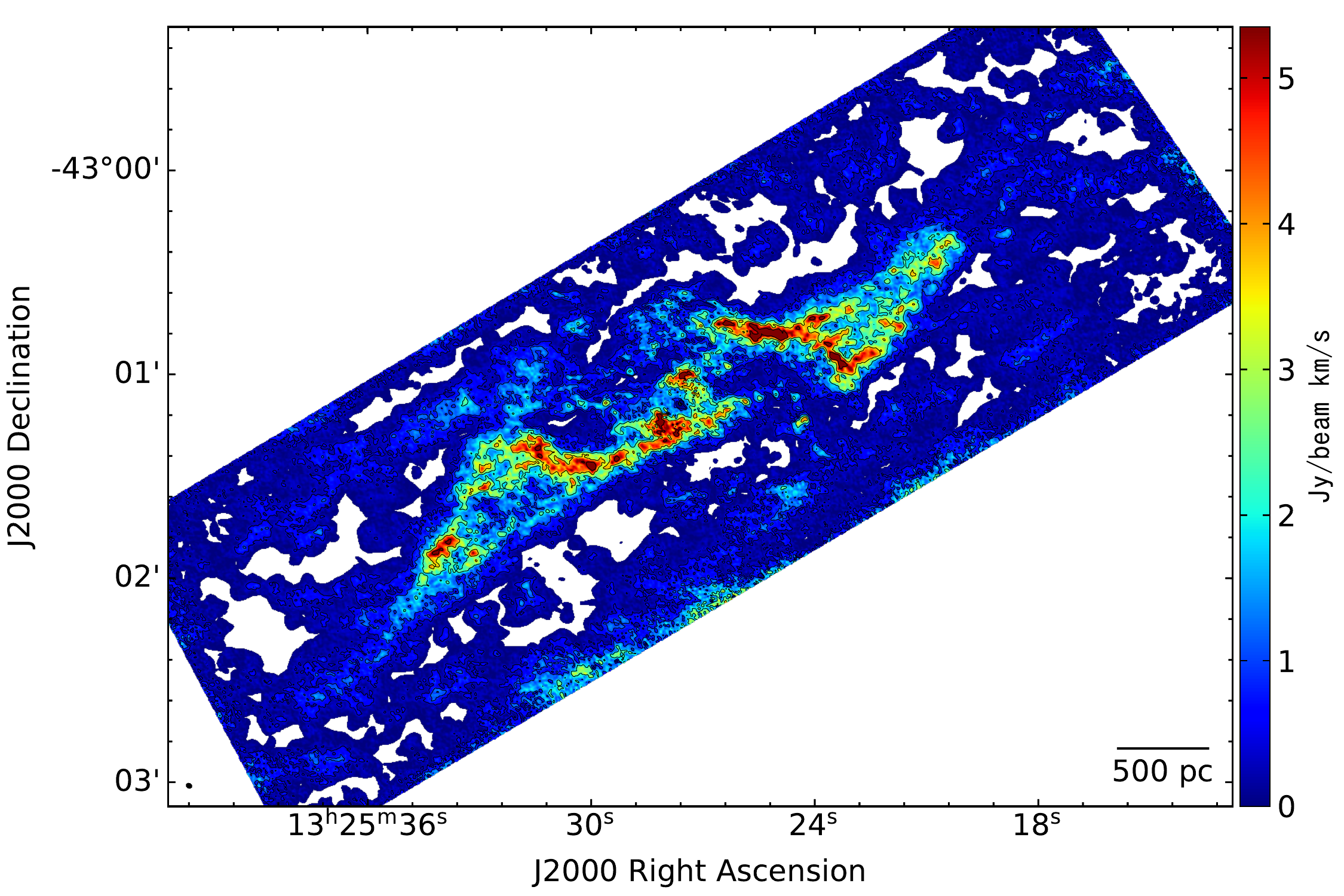}\\
 \end{center}
 \caption{
  CO(1--0) integrated intensity map of the molecular disk of Cen~A. Contour levels are at $3, 7, 15, 25$ and $40\,\sigma$, where $\sigma=0.14$\,Jy\,beam$^{-1}$\kms.
 The synthesized beam is shown as a filled red ellipse at the left bottom corner. {The white regions inside the map were masked (see \S \ref{obs} and Paper\,{\sc i} for details).}
 \label{mom}}
 \end{figure*}

 Thanks to ALMA's high angular resolution, sensitivity as well as dynamic range, we were able to resolve the molecular component into tens of parsec scale clouds.
 The CO(1--0) integrated intensity map of the inner molecular component of Cen~A is shown in Fig.\,\ref{mom}, {obtained as 
explained in Paper\,{\sc i} by smoothing the CO(1--0) data cube to calculate masks that were later applied to the original data cube.} The velocity field and velocity width maps were also presented in Paper\,{\sc i}.

 \section{CO(1--0) GMC Catalog}
 \label{methods}


 We identified Giant Molecular Clouds (GMCs) with the CPROPS package \citep{2006PASP..118..590R} and derived cloud properties.
 The CPROPS algorithm searches for emission in connected discrete regions (so called {\it islands}) above $4\sigma$ and velocity width of 4\,\kms . These {\it islands} are extended to include all adjacent pairs of channels that have emission above $2\sigma$.
  { The parameters we use in CPROPS are {\sf THRESH}=4,  {\sf EDGE}=2. {The cloud decomposition was done by the CPROPS default setting.} In addition, we set the minimum peak of an island to 6\,$\sigma$ ({\sf MINPEAK}=6). }
  Meanwhile, it excludes other {\it islands} that does not fulfil the requirement of two spatial resolution elements (i.e. twice the synthesized beam of $\sim 1\arcsec$) as minimum projected area,  and/or have a low signal-to-noise ratio S/N $<$ 5$\sigma$ { in flux}.  
  We excluded  cloud candidates outside the primary beam response at a 60\,\% power level in order to minimize uncertainties due false detections at the edges of the field of view and primary beam correction.  {Also, the algorithm compares the moments of
 the emission to distinguish separated and combined clouds. If moments vary by more than a set
 fraction by combining the two clouds, they are categorized as distinct.
 This is controlled by parameters {\sf SIGDISCONT} and {\sf FSCALE}, and we use the defaults 1 and 2, respectively (i.e. $>$ 200\% flux variation in merging a cloud would be significant).}

 A total of 689 GMCs were identified by the algorithm and their properties are listed in Table\,\ref{tbl2}.
 The table presents the cloud id, the cloud position in relative coordinates (in arcsec) relative to the center position of the AGN at $\alpha=13^{\rm h}25^{\rm m}27\fs615$ $\delta=-43$\arcdeg01\arcmin08\farcs80 ($\Delta$R.A., $\Delta$Decl.), the GMC  mean velocity ($v_{\rm LSR}$), the velocity dispersion ($\sigma_V$), the size before beam deconvolution ($\sigma_{\rm maj}\times \sigma_{\rm min}$),
 the radius ($R$), the CO(1--0) flux density ($S_{\rm CO(1-0)}$), and the
 virial mass ($M_{\rm vir}$).
  The nomenclature and convention is as in \citet{2018ApJ...864..120M}, except that they used CO(2--1) instead of the CO(1--0) line.
  We use the CPROPS measurements, which are obtained by extrapolation of the emission profiles to the zero intensity level.
 The radius is calculated as
 $R=1.91\,\sqrt{[\sigma_{\rm major}^2-\sigma_{\rm beam}^2]^{1/2}[\sigma_{\rm minor}^2-\sigma_{\rm beam}^2]^{1/2}}$ 
 , where $\sigma_{\rm beam}$ is the synthesized beam size, and $\sigma_{\rm major}$ and $\sigma_{\rm minor}$ the extrapolated rms sizes of the GMC's major and minor axis.
  The virial mass is obtained using equation $M_{\rm vir}=189\,\Delta V^2\,R$ [\Msol], which assumes that clouds are spherical and in virial equilibrium, with a volume density profile described by a truncated power law  $\rho\propto r^{-1}$ \citep{1987ApJS...63..821S}.
 $\Delta V$ is the full width at half maximum (FWHM) velocity line width in \kms\ expressed as { $\Delta V = 2\sqrt{2\ln 2}\, \sigma_V$}.
 We note that usually the assumption for the cloud shape is often spherical with a uniform density gradient, but it is obvious that this is not always true in practice.
 However, virial masses are expected to depend weakly on cloud shape (within 10\% for a cloud aspect ratio difference of about an order of magnitude, \citealt{1992ApJ...395..140B}).
  As for changes due to different density profiles, the assumption of $r^{-1}$ is probably the most realistic, but if proportional to $r^{-2}$ the actual virial masses would only decrease by $\sim30$\,\% from the derived ones assuming $r^{-1}$ \citep{1988ApJ...333..821M}.

 The bootstrapping method (with 10000 repetitions) was used to derive the uncertainty of each parameter in CPROPS.
  We note that this uncertainty does not include the intrinsic error of the spatial and velocity resolution limits of the CO(1--0) data nor the CO flux measurements. However, we include these sources of uncertainty in the $X_{\rm CO}$ factor later discussed in \S\,\ref{xcofactor}.

 The CO(1--0) luminosity is given by $L^{\prime}_{\rm CO(1-0)} = (c^2/2k_B)\,S_{\rm CO(2-1)}\,\nu_{\rm obs}^{-2}\,D_{\rm L}^2$, or $L^{\prime}_{\rm CO(1-0)} = 3.25\times10^7\,S_{\rm CO(1-0)}\,\nu_{\rm obs}^{-2}\,D_{\rm L}^2$ [K\,\kms\,pc$^2$], being $c$ the light speed, $k_B$ the Boltzmann constant, $S_{\rm CO(1-0)}$  the integrated CO(1--0) line flux density in Jy\,\kms, $\nu_{\rm obs}$ is the observed frequency in GHz,  and $D_{\rm L}$ the luminosity distance to the source in Mpc \citep[][]{2005ARA&A..43..677S}.
 The luminosity mass of the clouds was calculated as $M_{\rm gas} =  4.3\,L^{\prime}_{\rm CO(1-0)}$ \citep[e.g.][]{2013ARA&A..51..207B}, where the 4.3 factor corresponds to an $I_{\rm CO}-N({\rm H}_2$) conversion factor of $X_{\rm CO}$ = 2 $\times$ $10^{20}$\,cm$^{-2}$\,{(K\,\kms)$^{-1}$}. This is the $X_{\rm CO}$ factor we use unless mentioned otherwise for the CND (see \S\,\ref{xcofactor}).


 \begin{figure*}
 \begin{center}
 \includegraphics[width=0.9\textwidth]{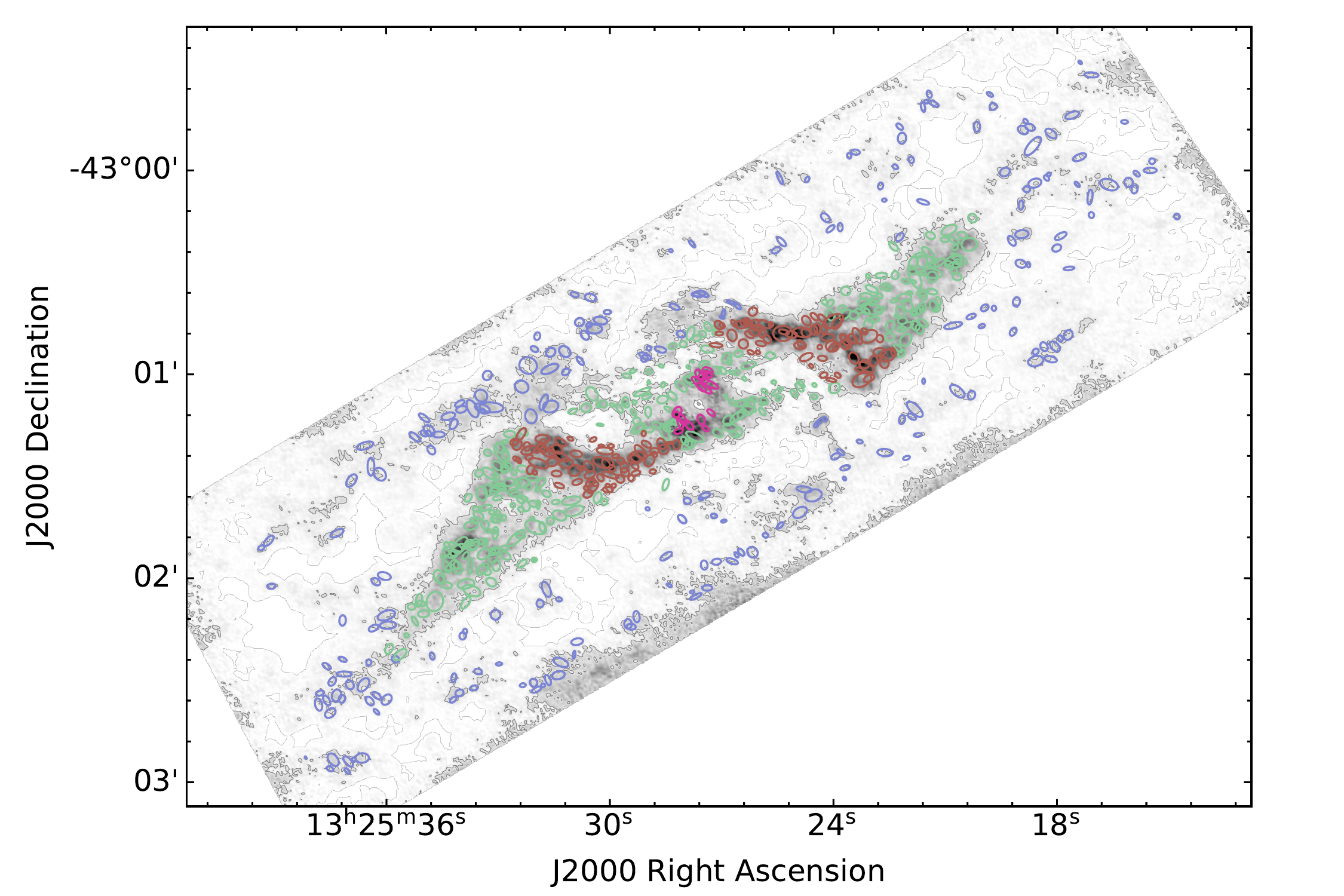}
 \end{center}
     \caption{The identified GMCs in Centaurus~A, overlaid on the CO integrated intensity image.  The gray contour is at a 5\,$\sigma$ level (0.7\,Jy\,\kms). The sizes of ellipses are equivalent to the effective major and minor radii of the GMCs. The four distinct regions are color coded.  The magenta, brown, green, and blue ellipses indicate the GMCs which are in the CND (central 220$\times400$\,pc, P.A.=155$^{\circ}$), the spiral arms \citep{2012ApJ...756L..10E}, the parallelogram structure region, 
  and the outskirts of the molecular disk region, respectively. We limit the identification of the GMCs in the CND to the velocity range of less than 534\,\kms\ or more than 564\,\kms\ (see \S\,\ref{methods} for details). \label{astrodendro}}
 \end{figure*}

 The GMC locations and sizes (effective major and minor diameters) are shown in Fig.\,\ref{astrodendro} overlaid on the integrated intensity CO(1--0) map. We separated the GMCs which belong to the CND, molecular arms, parallelogram structure, and the outermost disk, as explained in  Paper\,{\sc i}. { We have excluded the GMCs within a radius of 2\,\arcsec\ from the galaxy center with the velocity range between 534\,\kms\, -- 564\,\kms\  to avoid contamination due to cleaning residuals from the strong absorption lines towards the AGN \citep[e.g.][]{2010ApJ...720..666E}. The number of identified GMCs which fall in this category is only five.}
 A caveat in the identification procedure is that due to the warping of Cen~A's disk, different molecular components may appear along the line of sight \citep[e.g.][]{2010PASA...27..396Q}. However, our spectral resolution is good enough to be able to separate different GMCs along the line of sight.

 \begin{figure*}
 \begin{center}
 \includegraphics[width=0.7\textwidth]{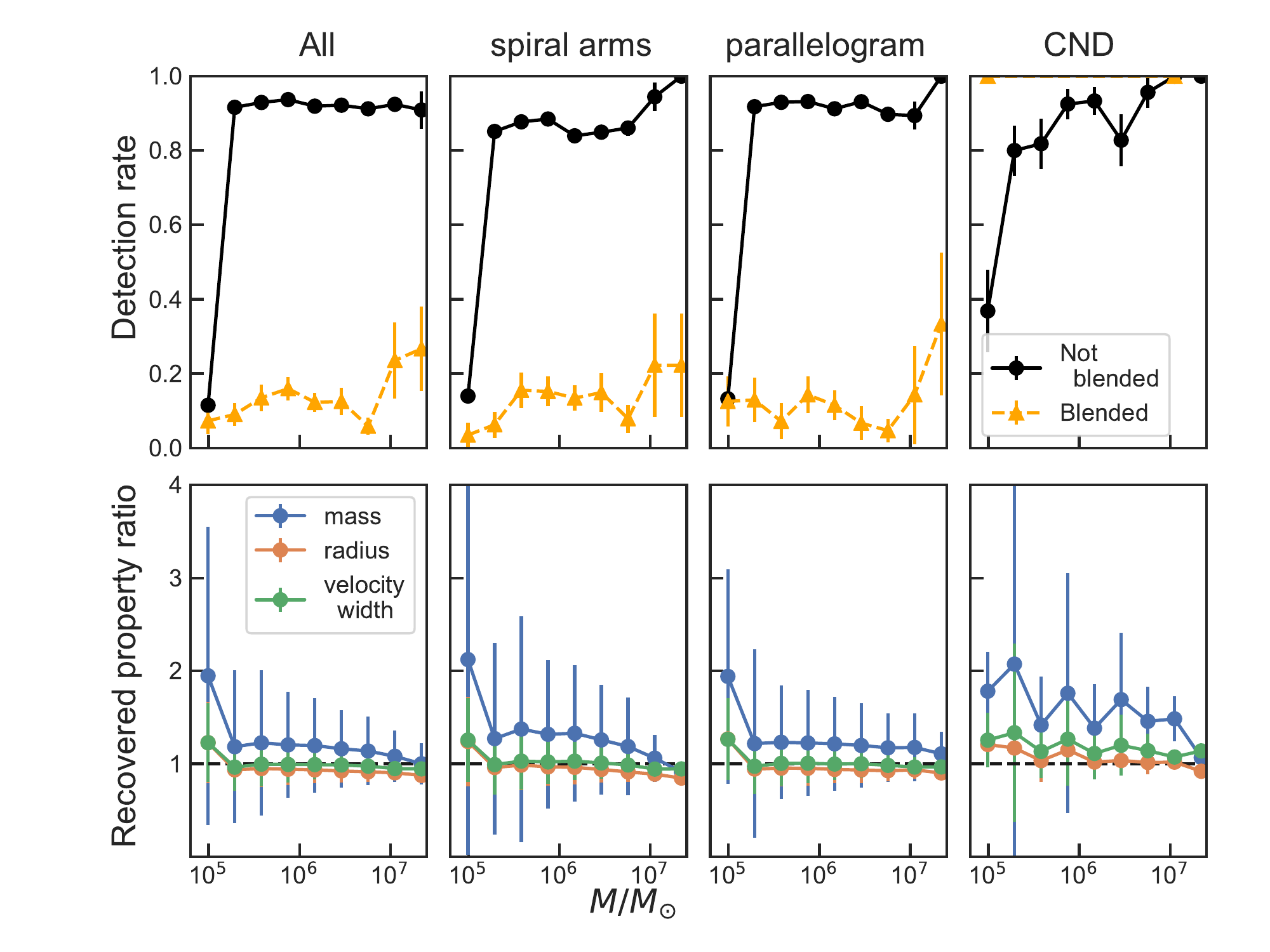}
 \caption{ {  Completeness limit assessment of our GMC survey using false source injection tests: ({\it Top panels}) The detection rate of the fake GMCs as a function of GMC mass. The (black) circles indicate the detection rate of fake GMCs that are well separated from any other pre-existing GMCs (not blended case), and the orange triangles the detection rate of fake GMCs that, although blended with pre-existing GMCs within their radii and velocity widths, are successfully identified and separated. ({\it Bottom}) The mean recovered mass, radius, and velocity width ratios of the not-blended fake GMCs as a function of GMC mass. 
 Columns from left to right show the plots for the entire region, spiral arms, parallelogram and CND regions. The error bars show the standard deviation. The horizontal dashed line indicates a recovered property ratio of unity.}
 \label{completeness}}
 \end{center}
 \end{figure*}

 Next we calculate the completeness limit of our GMC survey by performing false-source injection tests. This allows us to investigate the robustness of the obtained cloud properties and mass spectrum shapes.  This is necessary because we would be unable to distinguish clouds above a given limit based on the sensitivity estimate if the clouds were located in crowded regions of such as the spiral arms or the parallelogram structure. In other words, blending effects may effectively raise the completeness limit.

 In our tests, the masses of the fake GMCs range from $\log(M[M_\odot])=4.85$ to $7.50$, with a bin increment of 0.3.
 Once a mass is given, the velocity width $\sigma_v$ and radius $R$ are uniquely defined by the empirical scaling laws $M \propto \sigma_v^{4}$ and $M \propto R^{2}$ \citep{1987ApJ...319..730S}.
 The fake GMCs are placed in the original data cube assuming three dimensional Gaussian profiles.
  The locations of the fake GMCs are chosen randomly within the data cube.
 We generated in total 1800 individual fake GMCs per mass bin except for the two most massive bins ($\log(M[M_\odot])=7.0$ and 7.3), where we generated 200 and 50. This is because we focus on the lower mass end in order to probe the completeness limit. We then fed the simulated data cube into CPROPS with the same setting as we used for the original datacube (see \S\,\ref{methods}).
 
 A fake GMC is defined as ``recovered'' if a GMC is successfully identified as a new GMC (i.e. in addition to the already existing GMCs) in the data cube, within the synthesized beam and one velocity channel of its input location, or ``non-recovered'' otherwise.
 In Fig.\,\ref{completeness} (top panel) we present the detection rate of the fake GMCs as a function of cloud mass.

 The fake GMCs can be either well separated from any of the previously identified GMCs in our catalog (we call them ``not blended''; {circle symbols in Fig.\,\ref{completeness} {\it Top} and {\it Bottom panels}), or located close (within its radius and velocity width) to a previously identified GMC (``blended''; triangle symbols in Fig.\,\ref{completeness} {\it Top panels}).
 In our tests, the number of blended fake GMCs per mass bin is in the range 15 -- 160, or about 1\% -- 10\% of the total.

 We find that most clouds in the mass bin $\log(M[M_\odot]) = 5.3$ and above are recovered by CPROPS as far as the GMCs are well isolated, and the detection rate is overall above 90\,\%. In the $\log(M[M_\odot])=5.0$ bin the detection rate drops down to about 10\,\%, which means that we are largely incomplete in that regime.
  For fake GMCs that are located close (position and velocity) to any of the pre-existing GMCs in the data cube (i.e. blended case), the detection rate remains less than about 20\,\% in mass bins $\log(M[M_\odot])\lesssim 7.0$.

 In Fig.\,\ref{completeness} (bottom panels) we also present statistics of the ratio of the main properties (mass, radius, and velocity width) for the recovered fake GMCs in our experiment, in the not blended case. 
 We define the recovered property (mass, radius, or velocity width) ratio as the ratio between the derived property of the GMC as obtained by CPROPS and the original input of the fake GMC.
  All the property ratios are very close to unity at mass bins $\log(M[M_\odot])\geq 5.3$.

 We also separate in Fig.\,\ref{completeness} the detection rate and recovered property ratios by region, i.e. for the fake GMCs located in the spiral arms, the parallelogram region, and the CND, which are probably the most crowded regions (either physically or in projection) and where blending effects in the identification may be most severe. We find that the detection rate for $\log(M[M_\odot])\geq 5.3$ becomes only slightly worse ($>85\,\%$) in the spiral arm region compared to that of the entire area (or the parallelogram region).
 In the CND the detection rate for $\log(M[M_\odot])\geq 6.0$ is also $>85\,\%$, and we adopt this value as our completeness limit.  We note that the number of fake GMCs is more limited ($N$ = 239 in total) and there are less data points per mass bin.
 The recovered mass, radius and velocity width ratios as a function of mass for these three regions are in agreement with the trends observed for the entire area, i.e. they are very close to unity. 
 
 In summary, we adopt a conservative completeness limit of $\log(M[M_\odot]) = 6.0$, which ensures that the detection rate of the artificially-injected sources is above 85\% in the different regions, and that the GMC properties are recovered well when the GMCs are not blended. CPROPS has difficulties separating well the blended cases, especially at the low mass regime, but physically having multiple GMCs with a similar location and velocity may mean that they belong to the same complex. The most affected region by blending effects is probably the spiral arms, but compared to other regions the detection rate only decreases by about 10\%, and the recovered property ratio trends are comparable to the other regions.

 \section{Results}
 \label{results}

 \subsection{ GMC Properties}
 \label{mainGMCprop}

The velocity dispersion $\sigma_V$, radius $R$ (after beam deconvolution), and luminosity of the GMCs in the molecular disk of Cen~A spans 1--25\,\kms,  7--96\,pc, and 1.6 $\times$ 10$^4$ -- 6.9 $\times$ 10$^6$\,K\,\kms\,pc$^2$, respectively. The median $\sigma_V$ of all GMCs is  6.3\,\kms .  Excluding GMCs in the CND the median is 6.1\,\kms , and for those GMCs in the CND, the median is twice that value, 12.4\,\kms . The median radius and luminosity of all GMCs is 38\,pc and 2.5 $\times$ 10$^5$\,K\,\kms\,pc$^2$.

 Fig.\,\ref{PropertiesR} shows the variation of velocity dispersion and size in bins of 100\,pc with distance from the galaxy center. 
 We find that both quantities are remarkably flat, but the increase of the velocity dispersion is apparent in the inner few hundred\,pc.
 To derive distances we assumed a simple geometry where the CND can be characterized by a disk of 200\,pc radius, inclination of 60\arcdeg , and P.A. of 155\arcdeg \citep{2017ApJ...843..136E}, while for larger radii the molecular disk has an (averaged) inclination of 80\arcdeg and a P.A. of 120\arcdeg \citep{2010PASA...27..396Q}. { Note that we did not correct the GMC properties for inclination.}

 \begin{figure}
 \begin{center}
 \includegraphics[width=0.5\textwidth]{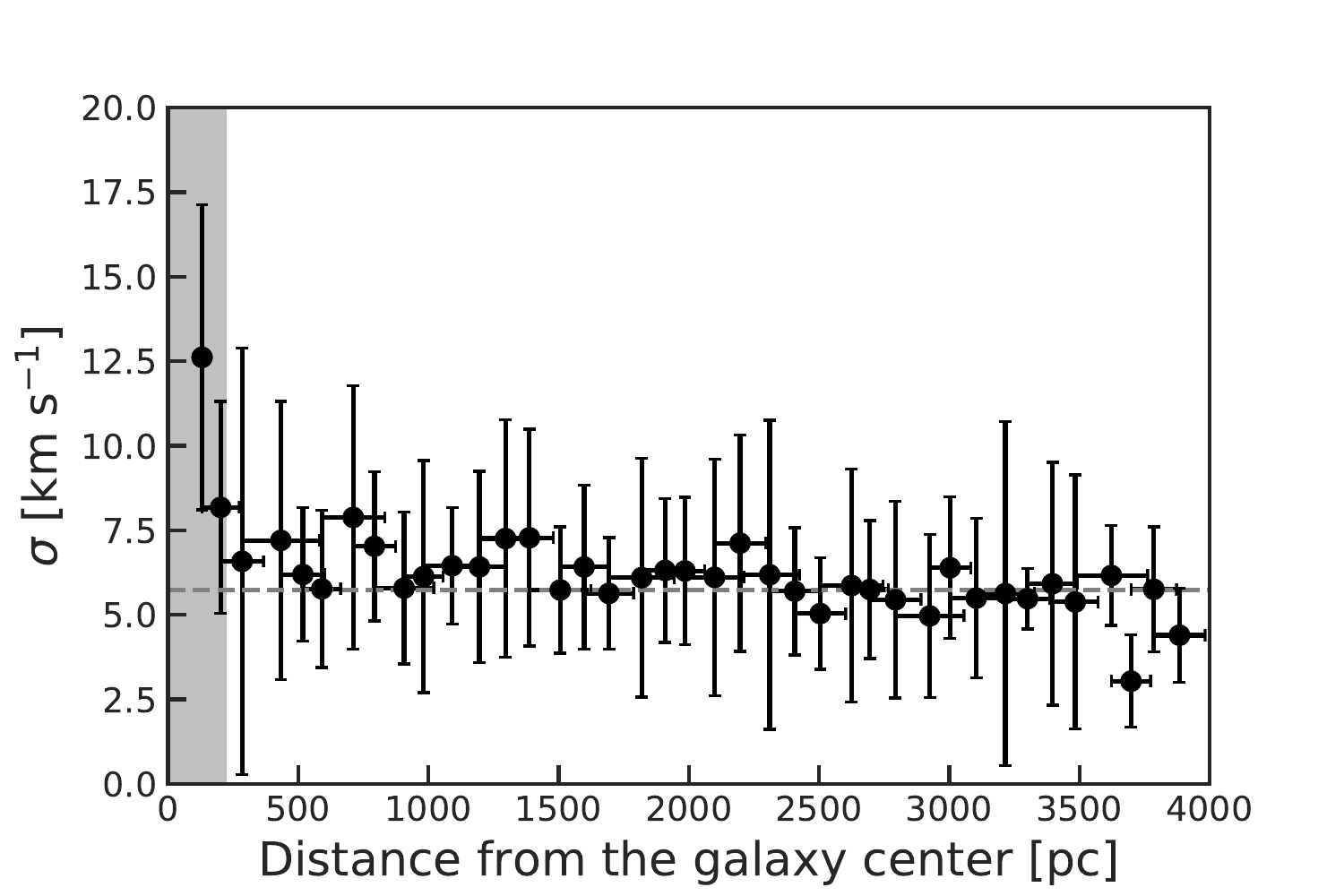}
 \includegraphics[width=0.5\textwidth]{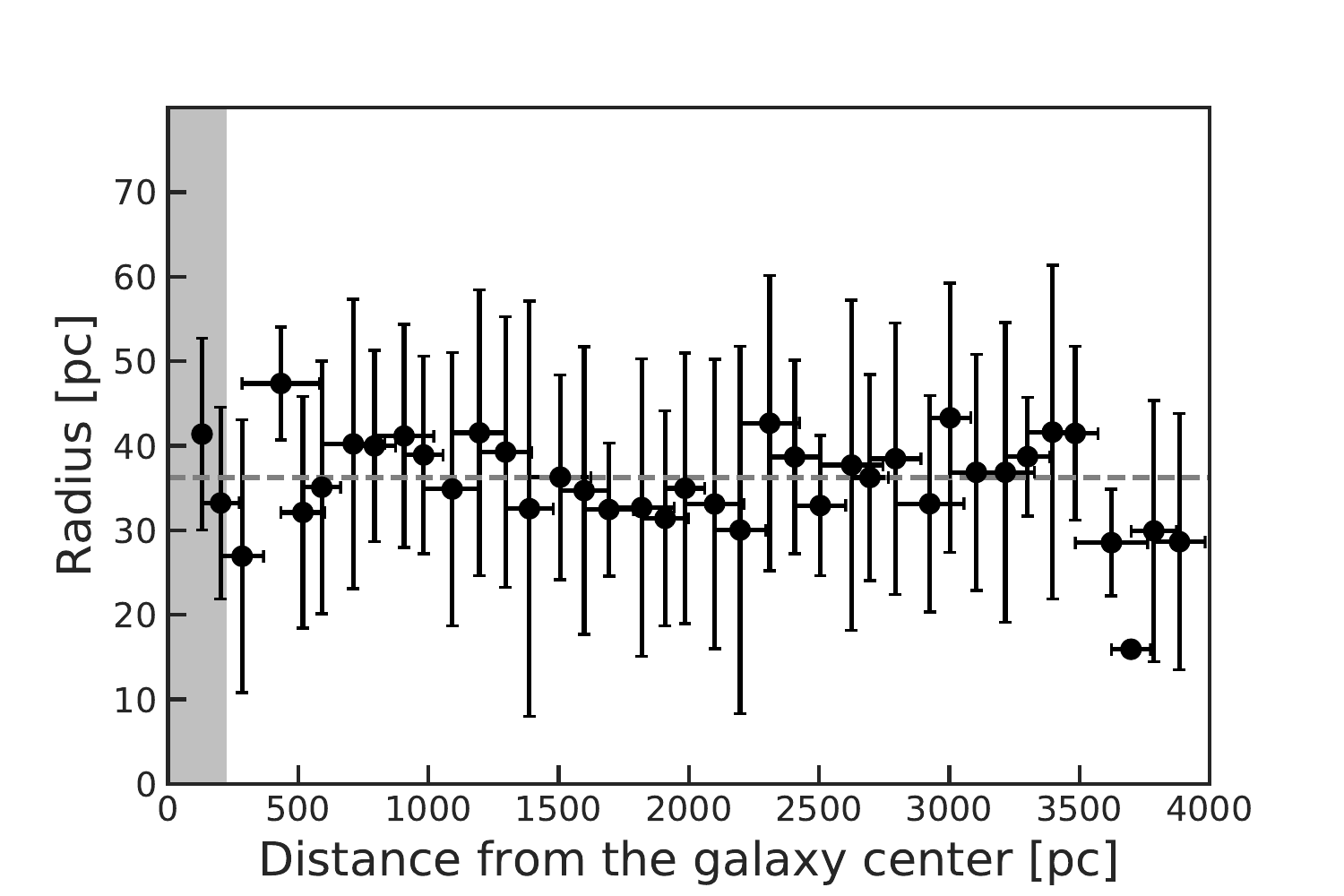}

 \caption{ The variation of velocity dispersion and radius as a function of the distance from the galaxy center. {Each data point is the median derived for the GMCs in every 100\,pc radial bin.} The error bars are the standard deviation of the values for the GMCs in that bin.  {The dashed lines show the median values for all GMCs.} The shaded area corresponds to the CND region ($\sim 200$\,pc radius). See \S\,\ref{mainGMCprop} for the simple assumptions made for the geometry of the disk in order to calculate distances.
 \label{PropertiesR}}
 \end{center}
 \end{figure}
 
 Table\,\ref{tbl4} presents the number of identified GMCs in each region  and the total CO luminosities compared to those obtained for the GMCs, with and without the extrapolation in CPROPS. 
 For the CO(1--0) luminosity of all the GMCs, we obtain $2.63\times10^8$\,K\,\kms\,pc$^2$. Therefore, 76\,\% of the total CO(1--0) luminosity from the molecular disk ($3.47\times10^8$\,K\,\kms\,pc$^2$) arises from molecular gas in GMCs with masses above 10$^5$\,$M_\odot$.
 The remaining 24\%  of the CO(1--0) luminosity mostly comes from the outer region (less than half of the CO luminosity there is recovered in GMCs) and parallelogram structure, likely in the form of smaller GMCs ($<10^5\,M_{\odot}$).

 There is a trend in the sense that the percentage of molecular gas arising from GMCs with masses larger than 10$^5$\,$M_\odot$ is smaller with increasing radius. For example, in the spiral arms (inner regions), most of the CO(1--0) emission arises from identified GMCs (81\% even without extrapolation), while in the outer disk most of the gas is in a diffuse component or in low-mass ($<$10$^5$\,$M_\odot$) GMCs.
 We note that the extrapolated CO luminosity of the GMCs in the spiral arms exceeds the total CO luminosity directly obtained from the CO(1--0) map. This means that the extrapolation below 2\,$\sigma$ down to zero-intensity in CPROPS results in an overestimation of the individual luminosities. This might be because in the spiral arms the GMCs are too crowded (spatially and in velocity), and thus the extrapolation becomes uncertain { (see also Fig.\,\ref{completeness}).
 The percentage of molecular gas in the form of GMCs in the outer regions is about half of the total CO luminosity.
 This means that more than half of the CO luminosity in the outer regions may arise from smaller GMCs ($<10^5\,M_{\odot}$).

 In the case of the CND region, this radial trend does not hold.
 The CO(1--0) luminosity in GMCs is 72\,\% and 33\,\% of the total CO(1--0) luminosity ($13.4\times10^6$\,K\,\kms\,pc$^2$) with and without extrapolation cases, respectively.
  This will be further discussed with the use of GMC mass spectra in \S\,\ref{massspectrum}.


 \subsection{Line Width - Size Scaling Relation} \label{larson-text}
 
 In this section we study the  line width - size scaling relation, which has been seen to hold at various scales and  measures the turbulent conditions of the molecular interstellar medium \citep[e.g.][]{1981MNRAS.194..809L}.
 It is generally seen that this relation increases as a power of radius $R$ [pc] such as in our Galaxy $\sigma_V=0.72\,R^{0.5}$\,\kms\, \citep{1987ApJ...319..730S,2009ApJ...699.1092H}.
 However,  quiescent molecular clouds and those in extreme environments such as starbursts present offsets with respect to each other of up to a factor of 10 in velocity dispersion \citep[e.g.][]{2001ApJ...562..348O}.
 
 The relation of these two parameters for the GMCs in the molecular disk of Cen\,A is plotted in Fig.\,\ref{larson}. 
 Next we compared with other works in the literature, including { the lenticular galaxy NGC\,4526 \citep{2015ApJ...803...16U} and  the spiral galaxy M51 \citep{2014ApJ...784....3C}}, where GMC identification and parameter calculation were carried out using CPROPS with a procedure similar to that presented here, and their corresponding datasets also have a similar resolution (20 -- 40\,pc, 2 -- 10\,\kms) and sensitivity ($2\times10^5-5\times10^5\,M_{\odot}$).
 Although there are many studies in the literature reporting GMC properties obtained using CPROPS as the choice of decomposition algorithm \citep[e.g.][]{2001ApJ...551..852H,2009ApJ...699.1092H,2001ApJ...562..348O,2011ApJS..197...16W,2008ApJ...686..948B,2013ApJ...772..107D,2015ApJ...801...25L,2018ApJ...864..120M,2018PASJ...70...73H}, we limited the comparison to the two datasets above since the derived cloud properties may be a strong function of the limiting spatial and spectral resolutions, as well as the sensitivity of the input data and decomposition parameters \citep[e.g.][]{2013ApJ...779...46H,2016ApJ...831...16L}.

 We find that the velocity dispersions of GMCs in the molecular disk of Cen~A are offset from the standard line width - size relation for the Milky Way disk \citep[indicated as a dashed line,][]{1987ApJ...319..730S}. The offset is 0.14\,dex from the standard relation (see Fig.\,\ref{larson}). { The GMCs identified in the CND are located at projected separations from the center of 5 to 11\arcsec, where the rotation curve flattens (e.g.  Fig. 9 in \citealt{2017ApJ...843..136E}), so we do not expect that the galactic rotation contribution to the velocity dispersion is large compared to further out in the galaxy disk.}
 We also plot the GMCs within the CND with a different symbol to show the clouds possibly affected by the extreme environments in the central regions of Cen\,A.
 The GMCs near the CND region tend to have larger velocity widths for a given radius than the rest, and the offset is 0.43\,dex higher than the standard Galactic disk line width - size relation, {or $\sim$ 0.3\,dex higher than those in the GMCs in the molecular disk}. We note that they are also offset from Galactic Center clouds \citep[dot-dashed line in Fig.\,\ref{larson},][]{2001ApJ...562..348O}.
 GMCs with large velocity widths are also reported in the centers of galaxies such as M\,51 and M\,83 \citep{2014ApJ...784....3C,2018PASJ...70...73H}.
 Within the GMCs of the molecular disk of Cen\,A we have not found regional variations in this scaling relation, except in the CND region.

 There is some level of correlation between the velocity dispersion and the radius of the GMCs in the molecular disk of Cen~A, with a correlation coefficient $\rho = 0.43$, (0.46 when excluding the CND clouds). This is in contrast with the lack of correlation in the early type (lenticular) galaxy NGC\,4526 for $\sim$100 resolved ($\sim$ 20\,pc) GMCs \citep{2015ApJ...803...16U}, with a correlation coefficient of $\rho = -0.15$. 

 \subsection{$I_{\rm CO}-N({\rm H}_2$) Conversion Factor {and Virial Parameters} in Cen~A} \label{xcofactor}

 We report for the first time in this object the $I_{\rm CO}-N({\rm H}_2$) conversion factor ($X_{\rm CO}$ factor) obtained using the virial method.
 In this method the CO luminosities and the virial masses of the clouds are compared to derive the $X_{\rm CO}$ factor \citep{2007ARA&A..45..565M,2013ARA&A..51..207B}.

 \begin{figure*}
 \begin{center}
 \includegraphics[width=0.8\textwidth]{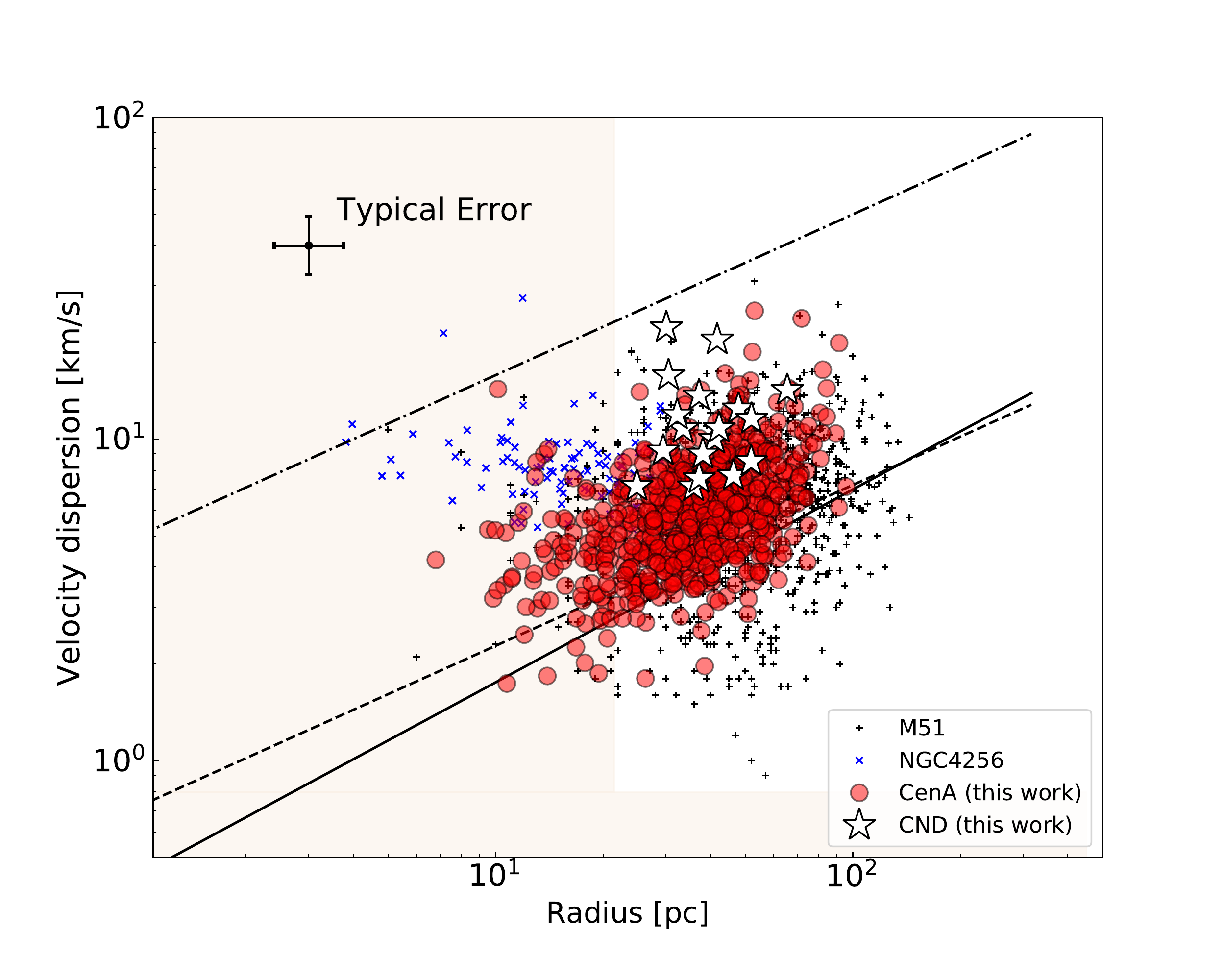}
 \end{center}
 \caption{Line width-size scaling relation for the identified GMCs in the molecular disk of Centaurus~A (CND: white stars; other regions: red circles), compared to the GMCs in {
 the lenticular galaxy NGC\,4526 \citep{2015ApJ...803...16U} and the spiral galaxy M51 \citep{2014ApJ...784....3C}}.
  The dashed, dot-dashed, and solid lines indicate the correlations found for the Galactic disk \citep{1987ApJ...319..730S}, Galactic center \citep{2001ApJ...562..348O}, and other galaxies \citep{2008ApJ...686..948B}.
 The limits of the shaded areas indicate the equivalent full width at half maximum (FWHM) of the synthesized beam and the equivalent velocity dispersion of a channel.
 \label{larson}}
 \end{figure*}

 \begin{figure*}
 \begin{center}
 \includegraphics[width={0.7\textwidth}]{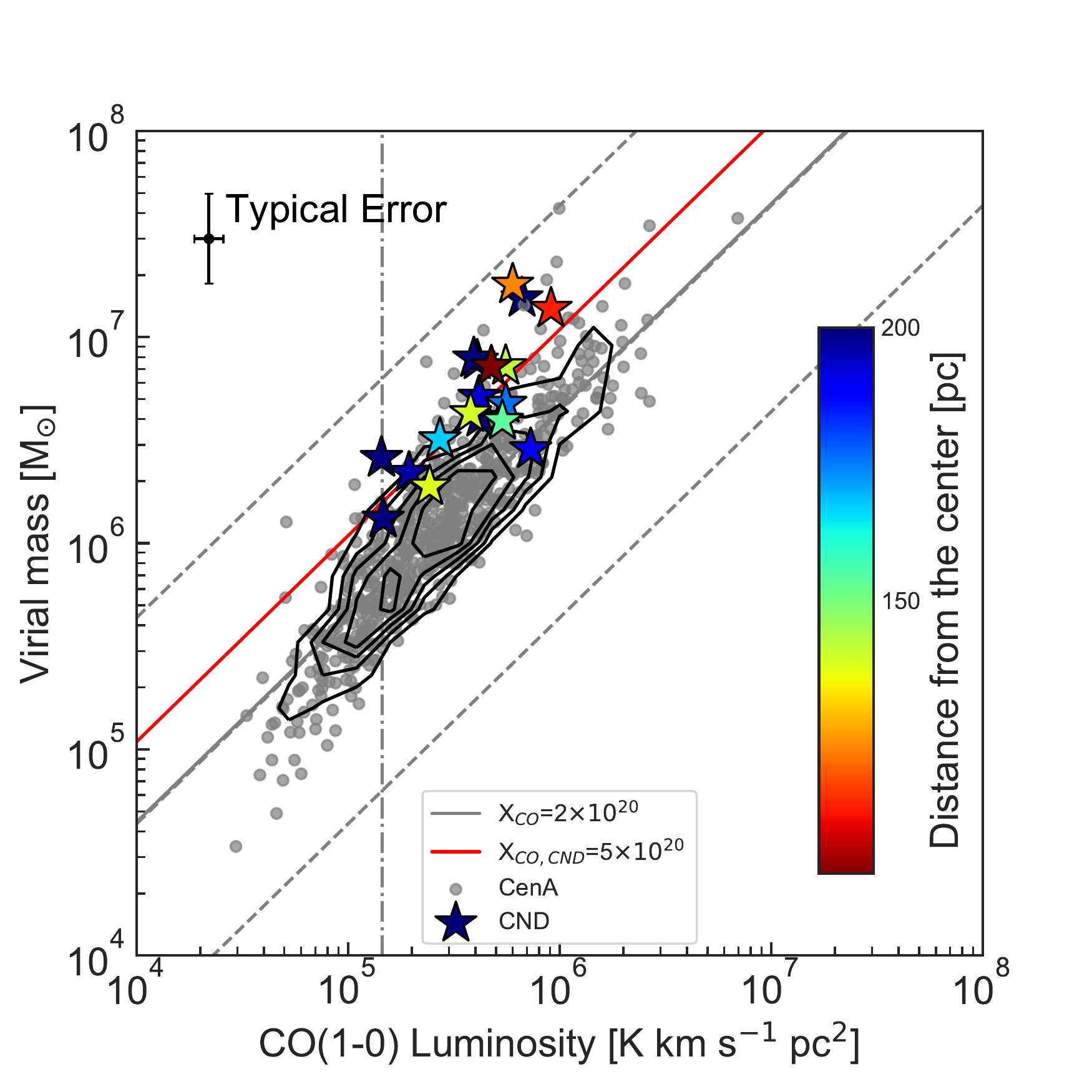}
 \end{center}
 \caption{
 CO luminosity - virial mass relation for the GMCs in the molecular disk of Centaurus A.
 The star symbols indicate the GMCs in the CND.
 The solid lines indicate the best linear fits (red color for the GMCs in the CND). The vertical line indicates the corresponding completeness limit. The dashed lines indicate $X_{\rm CO}$=0.2, 2, 20$\times10^{20}$\,cm$^{-2}$\,{(K\,\kms)$^{-1}$} for reference. The typical error in both coordinates is shown at the top left side of the plot.
  {The distances from the center of the individual GMCs in the CND region (see \S\,\ref{xcofactor}) are indicated with a color scale from 100 to 200\,pc.
 The contours show the number density of GMCs in this plot, and the level spans from 5 to 25 independent data points per 0.16\,dex $\times$ 0.16\,dex cell, in bins of 5.}
 \label{xfactor}}
 \end{figure*}

 Fig.\,\ref{xfactor} shows the tight correlation between virial masses and CO(1--0) luminosities for the identified GMCs (with a correlation coefficient of $0.72$).
  The best fit slope is 4.4 $\pm$ 2.0, which yields a conversion factor of $X_{\rm CO}$ = ($2 \pm 1$)$\times10^{20}$\,cm$^{-2}$\,{(K\,\kms)$^{-1}$}.
 For the estimation of the uncertainty of the $X_{\rm CO}$ factor, we assumed a 5\% gain uncertainty in the absolute amplitude calibration of the CO data (quoted from the ALMA Proposer's Guide and confirmed with the observed calibrators), and about 40--50\% uncertainty in the cloud property measurements (i.e. luminosity and virial masses), as well as the fitting error itself.
 This conversion factor is similar within the error bar to the standard Milky Way disk value \citep[$X_{\rm CO}$=2$\times10^{20}$\,cm$^{-2}$\,{(K\,\kms)$^{-1}$},][]{1987ApJ...319..730S,1988A&A...207....1S,2009ApJ...699.1092H,2013ARA&A..51..207B}.

 There is a certain amount of scatter in Fig.\,\ref{xfactor}. {No regional variations in this scaling relation are found, except in the CND. The data points for the CND are preferentially found toward higher $X_{\rm CO}$ factor values. There are some outliers at low (high) $X_{\rm CO}$ factor regimes, but these are mostly in the external (inner) regions of the CND. We used a color scale in Fig.\,\ref{xfactor} to represent the distance of each GMC to the center.

  The resulting conversion factor for the CND is larger than that of the molecular disk, $X_{\rm CO} = (5 \pm 2) \times10^{20}$\,cm$^{-2}$\,(K\,\kms)$^{-1}$. The total molecular mass of all the GMCs in the CND  is $8.6\times10^7$\,\Msol\ using $X_{\rm CO} = 5 \times10^{20}$\,cm$^{-2}$\,{(K\,\kms)}$^{-1}$, or $3.5\times10^7$\,\Msol\ if we use a constant $X_{\rm CO} = 2 \times 10^{20}$\,cm$^{-2}$\,(K\,\kms)$^{-1}$. The former value is consistent with the previously obtained value by \citet{2017A&A...599A..53I}, (9.1$\pm$0.9)$\times10^7$\,\Msol.}

 Fig.\,\ref{XcoR} shows the variation of $X_{\rm CO}$ in bins of 100\,pc as a function of the distance from the galaxy center. 
  We find that the  $X_{\rm CO}$ factor is flat at $X_{\rm CO}=2\times$ 10$^{20}$\,cm$^{-2}$\,(K\,\kms)$^{-1}$ for radii  $>200$\,pc, but we see a tentative trend where $X_{\rm CO}$ gradually increases toward the galaxy center to values $\sim(2-3)\times$ larger, which is then translated into an increase in the luminosity mass and gas surface density. Note that the minimum number of GMCs per bin within the molecular disk occurs in the CND, but it is still around 10.

 \begin{figure}
 \begin{center}
 \includegraphics[width=0.5\textwidth]{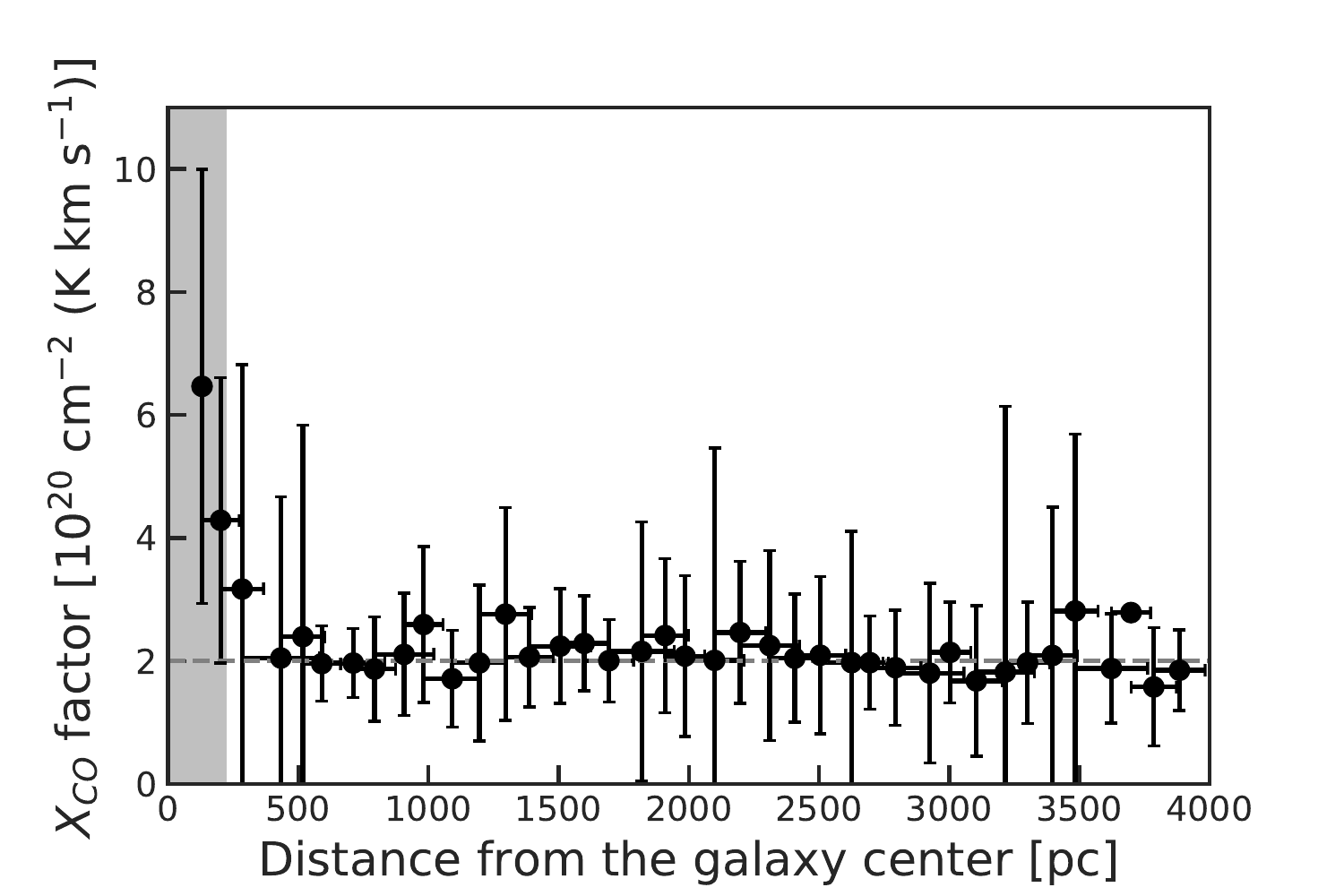}
 \includegraphics[width=0.5\textwidth]{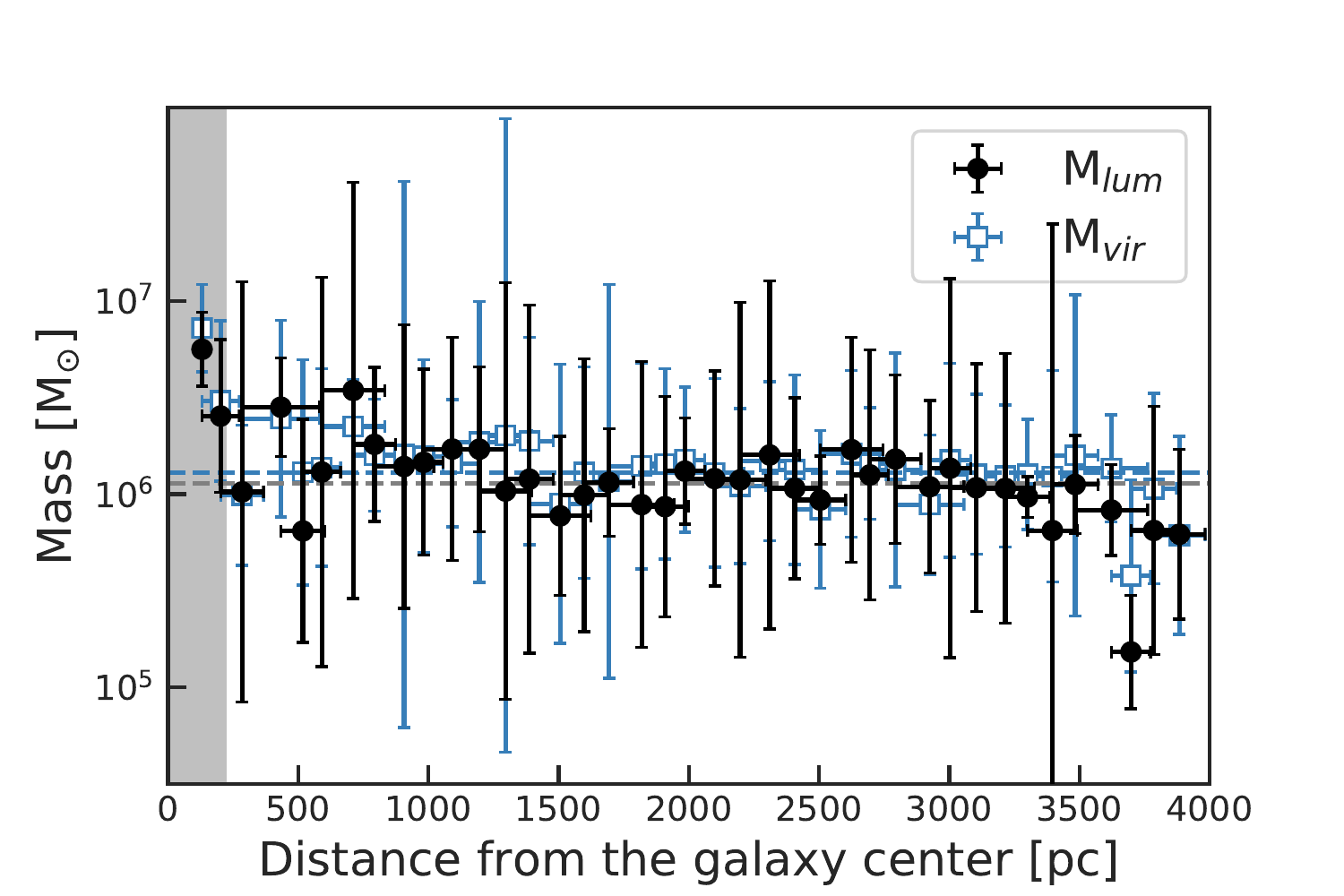}
 \includegraphics[width=0.5\textwidth]{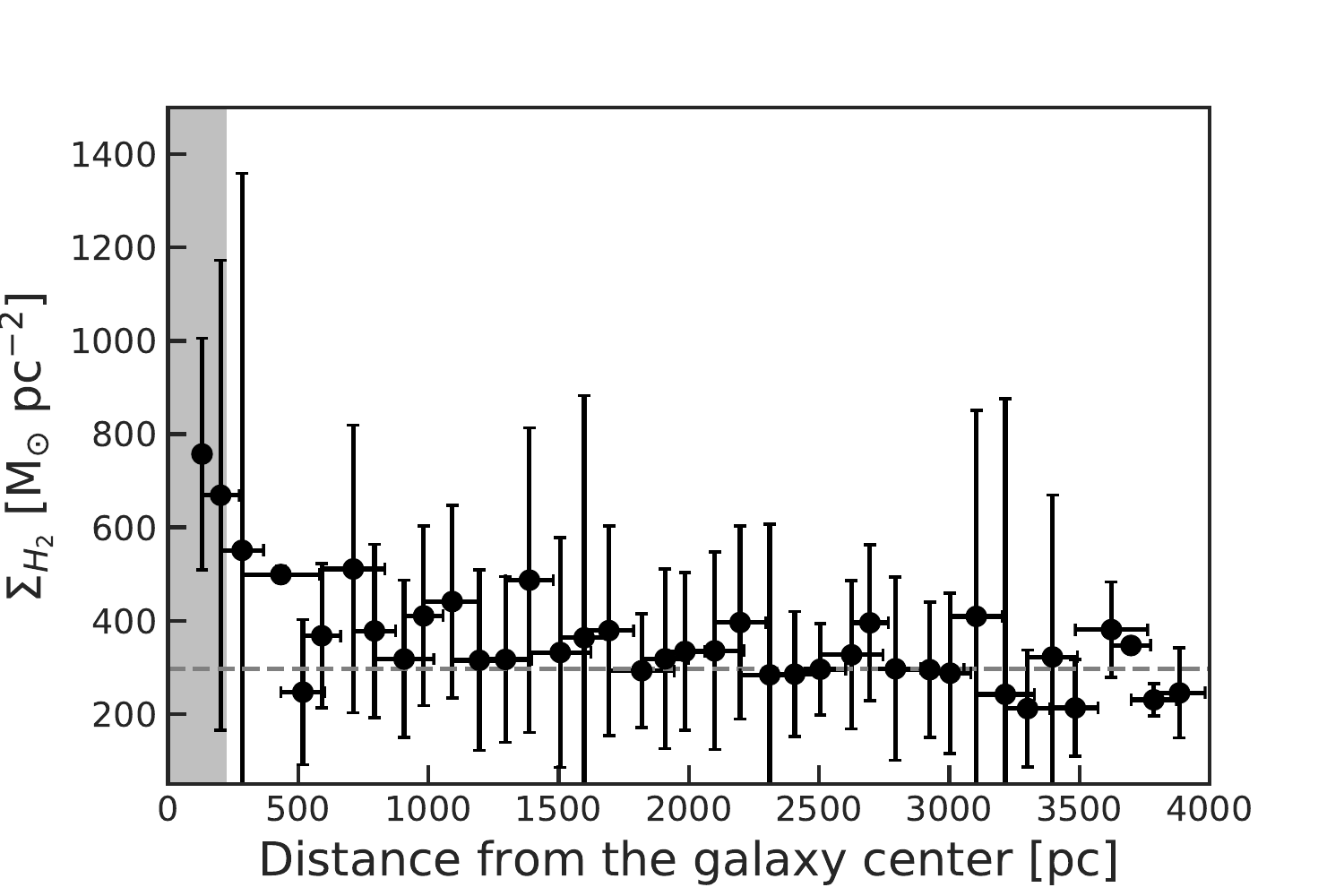}

 \caption{ The  $X_{\rm CO}$ factor, luminosity and gas surface density as a function of the distance from the galaxy center.{Each data point is the median of the $X_{\rm CO}$ factors derived for GMCs in every 100\,pc radial bin.} The error bars are the standard deviation of the values for the GMCs in that bin. {The dashed lines show the median values for all GMCs.} The shaded area corresponds to the CND region ($\sim 200$\,pc radius). See \S\,\ref{mainGMCprop} for the simple assumptions made for the geometry of the disk in order to calculate distances.
 \label{XcoR}}
 \end{center}
 \end{figure}

  { Fig.\,\ref{virial} shows the histogram of the virial parameters of the identified GMCs in Cen~A.  The median value of $\alpha_{\rm vir}$ is 1.0 (the standard deviation is 0.8), as expected because the used $X_{\rm CO}$ factor is the same as that we obtain for the molecular disk of Cen~A. We do not find any regional variation in the virial parameters. For the GMCs in the CND, if we use a common  $X_{\rm CO}$ factor, 
 we find values that are slightly offset, with a median value of $\alpha_{\rm vir}$ = 2.8  (the standard deviation is 1.5).
}
 
  \begin{figure}
 \begin{center}
 \includegraphics[width=0.5\textwidth]{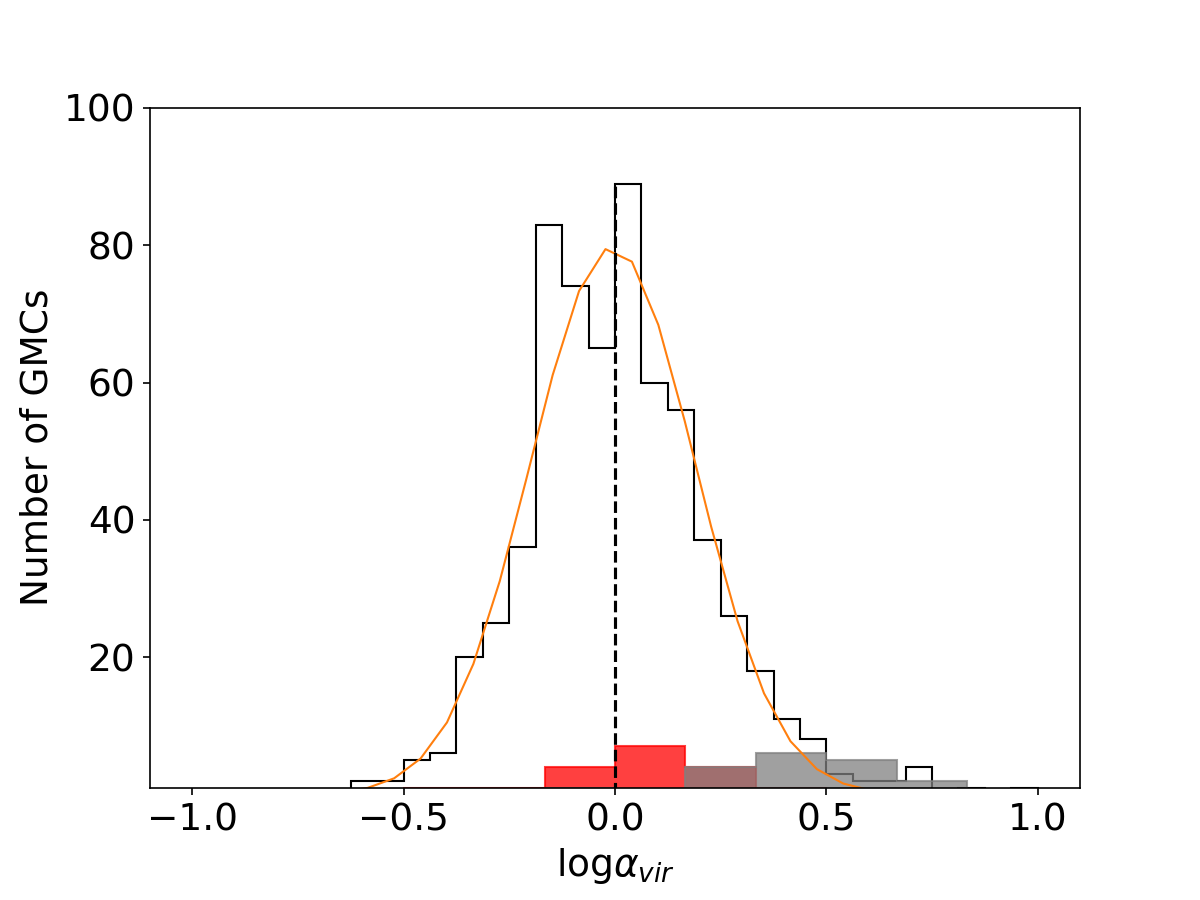}
 \end{center}
 \caption{The virial parameter ($\alpha_{\rm vir}$) distribution for the population of GMCs in the molecular disk of Centaurus~A, using the conversion factor $X_{\rm CO}$ = $2\times10^{20}$\,cm$^{-2}$(K\,\kms)$^{-1}$. The solid line is a Gaussian fit. The vertical (dashed) line indicates the median value for all GMCs, $\alpha_{\rm vir}=1.0$.
 The red and gray filled histograms are the $\alpha_{\rm vir}$ distributions for the GMCs in the CND using $X_{\rm CO}$ = $5\times10^{20}$\,cm$^{-2}$(K\,\kms)$^{-1}$ and $X_{\rm CO}$ = $2\times10^{20}$\,cm$^{-2}$(K\,\kms)$^{-1}$, respectively.
 \label{virial}}
 \end{figure}

 \begin{figure*}
 \begin{center}
 \includegraphics[width=0.8\textwidth]{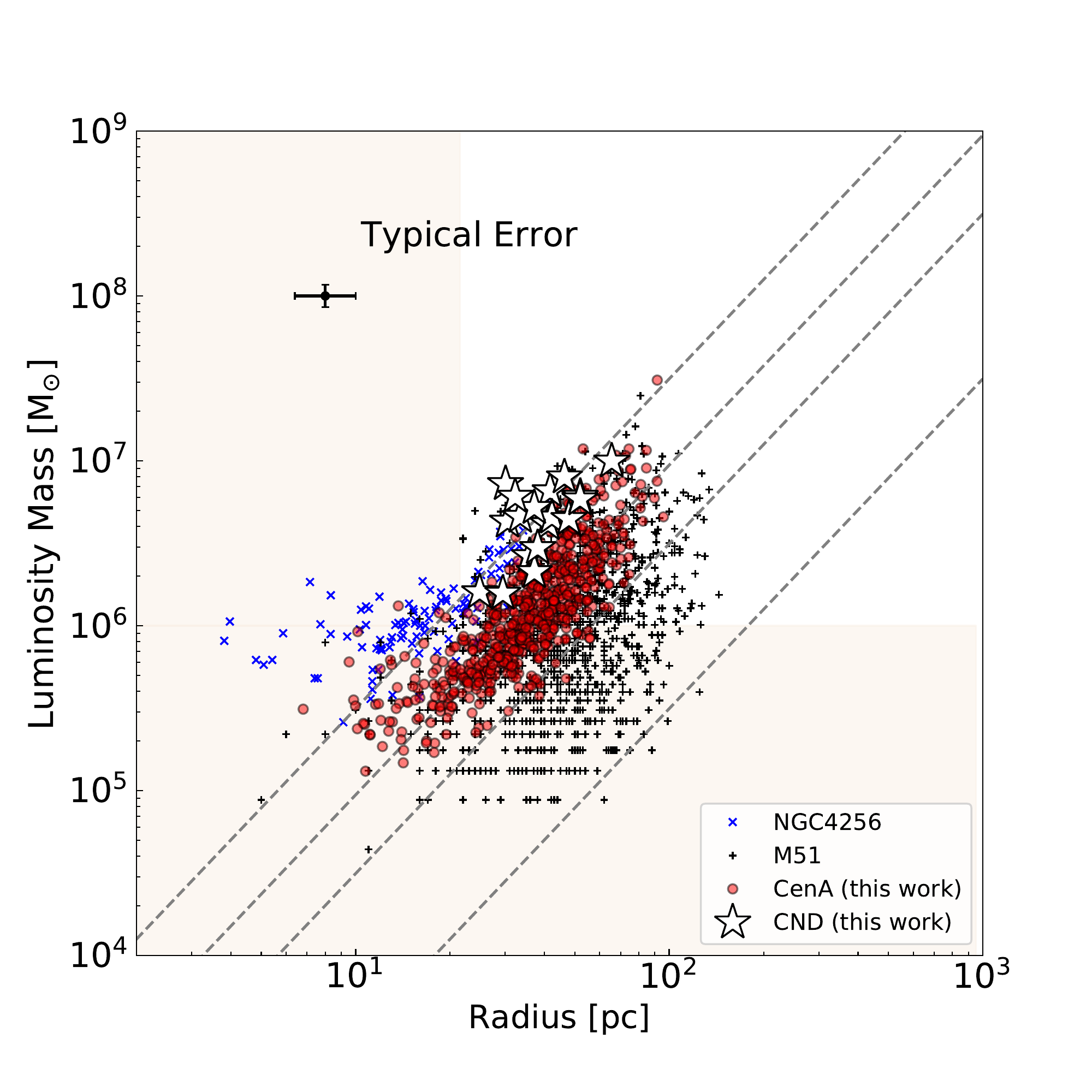}
 \caption{
 CO(1--0) luminosity mass as a function of cloud radius for the identified GMCs in the molecular disk of Centaurus~A. The dashed lines indicate 10, 100, 300 and $10^3$\,$M_{\odot}\,$pc$^{-2}$.
 The white stars represent the GMCs in the CND, using $X_{\rm CO}$ = $5\times10^{20}$\,cm$^{-2}$(K\,\kms)$^{-1}$. For other regions of the molecular disk of Centaurus~A we use $X_{\rm CO}$ = $2\times10^{20}$\,cm$^{-2}$(K\,\kms)$^{-1}$. Other symbols are as in Fig.\,\ref{larson}. The limits of the shaded areas indicate the equivalent FWHM of the synthesized beam in the x-axis and the completeness mass limit in the y-axis.
 \label{massR}}
 \end{center}
 \end{figure*}

 \section{Discussion}
 \label{discussion}

 \subsection{ The Large $X_{\rm CO}$ Factor in the CND}

 {
 The conversion factor obtained for the CND, $X_{\rm CO} = (5 \pm 2) \times10^{20}$\,cm$^{-2}$\,(K\,\kms)$^{-1}$,  is large compared to that in other regions in the molecular disk of Cen~A and other galaxy centers.
 Larger values for the CND were already reported by \citet{2014A&A...562A..96I} using an independent method based on modeling with Large Velocity Gradient (LVG) analysis of the CO spectral line energy distribution globally toward the CND. The value Israel et al. found is $X_{\rm CO}$ = 4 $\times10^{20}$\,cm$^{-2}$\,(K\,\kms)$^{-1}$ with an uncertainty of a factor of 2, which is consistent with our result.

A possible reason for this larger value than the Galactic $X_{\rm CO}$ is low metallicity conditions. In the low metallicity regime, examples are dwarf galaxies such as LMC \citep[$X_{\rm CO} \simeq 4 \times$ 10$^{20}$\,cm$^{-2}$\,(K\,\kms)$^{-1}$,][]{2008ApJS..178...56F} and the SB dwarf galaxy NGC\,5253 \citep[$X_{\rm CO}=4 \times$ 10$^{20}$\,cm$^{-2}$\,(K\,\kms)$^{-1}$,][]{2018ApJ...864..120M}. { The $X_{\rm CO}$ obtained by virial mass analysis depends on the physical resolution of the observations \citep[e.g.][]{2013ARA&A..51..207B}, in the sense that studies with finer spatial resolution systematically return lower $X_{\rm CO}$ than coarser resolution studies. The resolution obtained in dwarf galaxy observations is usually better, so the $X_{\rm CO}$ factor would be even higher when scaled to our resolution. }
 At any rate, the increase of the conversion factor in the CND is probably not due to lower metallicity conditions because \citet{2017A&A...599A..53I} showed that the metallicity in the disk of Cen~A is relatively constant 0.7 - 0.8 Z$_\odot$ both in the CND and in the outer disk regions.

  The large $X_{\rm CO}$ factor toward the CND of Cen\,A is a remarkable result because it is the opposite trend to that seen in the central parts of galaxies  and in molecule-rich SBs such as mergers where $X_{\rm CO}$ is often depressed \citep{2013ARA&A..51..207B,2013ApJ...777....5S}.
 The $X_{\rm CO}$ factor in galaxy centers including the Galactic Central Molecular Zone appear to be 3--10 times lower than the Galactic disk conversion factor, which is likely due to a combination of lower opacities partly because of larger line widths \citep[e.g.][]{1993A&A...274..148G,1995ApJ...452..262S,1998A&A...331..959D,2001ApJ...551..687M,2001ApJ...562..348O,2003A&A...406..817I,2006A&A...445..907I,2009A&A...506..689I,2009A&A...493..525I,2011MNRAS.411.1409W,2012ApJ...751...10P}}.
  Although the large line width condition may also apply to the CND of Cen\,A, still we find a larger $X_{\rm CO}$ factor. This can be due to a combination of higher excitation conditions together with the existence of molecular gas that is CO-dark \citep[e.g.][]{2012ApJ...751...10P,2018MNRAS.478.1716P}.
 Since the radiation field due to SF in the CND is expected to be very low \citep[][Paper {\sc I}]{2017A&A...599A..53I} the reason for the lower CO abundances may be the energetic radiation and cosmic rays from the AGN.
 We can exclude the possibility that it is simply due to a resolution effect because other studies of centres of galaxies and SBs were observed with coarser resolution, so the $X_{\rm CO}$ factor would be even lower when scaled to our resolution.

  A consequence of the larger $X_{\rm CO}$ factor in the CND is that we confirm the large average gas-to-dust mass ratio when compared to the outer disk as found by \citet{2012MNRAS.422.2291P} and \citet{2014A&A...562A..96I}. \citet{2012MNRAS.422.2291P} obtained a gas-to-dust mass ratio of 275 for the CND assuming the standard Milky Way $X_{\rm CO}$ factor, which would be further increased to 690 if we use our larger $X_{\rm CO}$ factor. Possible causes of the large gas-to-dust mass ratio might be dust sputtering by X-rays originating in the AGN or the removal of dust by the jets \citep{2012MNRAS.422.2291P}, although there are further uncertainties caused by the assumed dust properties \citep{2017A&A...599A..53I}.

 \subsection{Gas Pressure Balance}

 In Fig.\,\ref{massR} we present the CO(1--0) luminosity mass as a function of cloud radius for the GMCs in the molecular disk of  Cen~A. This is compared with other nearby galaxies and our Galaxy.
  Most GMCs in the molecular disk of  Cen~A are aligned along the line of surface density of $\Sigma_{\rm  H_2}$ $\simeq$ 300\,\Msol\,pc$^{-2}$ (the best fit is 315 $\pm$ 52\,\Msol\,pc$^{-2}$), higher than the general trend for the molecular clouds in our Galaxy and other nearby galaxies.
 The GMCs in the CND of Cen~A are aligned along the line of a surface density of $\Sigma_{\rm  H_2}$ $\simeq$ 10$^3$\,\Msol\,pc$^{-2}$, similarly to the GMCs in the lenticular NGC\,4256 \citep{2015ApJ...803...16U}.
  {The higher surface densities found with respect to other spiral galaxies for a given physical scale, together with the higher line widths, are likely related to differences in the environment.}

We probe the role of external pressure in confining molecular clouds with the relation of $\sigma_V^2/R$ and the gas mass surface density  \citep{2011MNRAS.416..710F} in Fig.\,\ref{pressure}.
 { The V-shaped curves in Fig.\,\ref{pressure} show the pressure-bound virial equilibrium solutions for six different external pressures. In short, the scaling coefficient $\sigma^2/R$ is given by:
 \begin{equation}
 \frac{\sigma^2}{R}\propto(C_1\Sigma + \frac{P_e}{\Sigma}),
 \end{equation}
 where $C_1$ is a constant, $P_e$ is the external pressure and $\Sigma$ is the gas surface density \citep[equivalent to Eq. 8 in ][]{2011MNRAS.416..710F}.
 Therefore, when $\Sigma$ is high compared to the external pressure, the gravitationally bound GMCs will be located in the plot along the straight line,
 but when the external pressure is high enough the GMCs will be located above it.
 }
 
 In Fig.\,\ref{pressure} we also see that the clouds close to the CND are characterized by relatively higher surface densities ($\Sigma_{\rm  H_2}$ $\geq10^3$\,\Msol\,pc$^{-2}$) and $\sigma_V^2/R$ ($\geq3$\,km$^2$\,s$^{-2}$\,pc$^{-1}$), compared to other GMCs in Cen\,A.
 Assuming the conversion factor for the GMCs in the CND as obtained from the virial method, the data points are seen to be clustered along the line of gravitationally-bound conditions  {(see top panel of Fig.\,\ref{pressure}) }.  If, on the other hand, we use the standard Milky Way disk conversion factor {(bottom panel of Fig.\,\ref{pressure}) , the GMC masses would be lower and then the data points move to the left} (i.e. lower surface densities) by a factor of $\sim2.5$, and external pressures of $P/k_B\sim10^6$ to $10^7$\,cm$^{-3}$\,K would  be needed to support the GMCs in the CND in addition to the self-gravity.
Other than in the CND, we have not found any evidence for potential regional variations in the relation $\sigma_V^2/R$ vs gas mass surface density.

{
\citet{2020ApJ...892..148S} compare the dynamical equilibrium pressure vs internal cloud pressure for GMCs in a sample of spiral galaxies. It is noted there that in environments with a large stellar content, such as in bulges, there might be higher dynamical equilibrium pressures. This will certainly be the case in Cen\,A. Also, we see that in Cen\,A, gas velocity dispersions are larger in these environments than in spirals, leading to higher dynamical equilibrium pressure as well. So overall we hypothesize that dynamical equilibrium pressures in the GMCs of Cen\,A will be higher in general to those in disk galaxies, especially in the central regions of the elliptical galaxy.
We see that in Cen\,A gas surface densities are typically larger than in spirals, which together with the larger velocity dispersion yields that internal cloud pressures would also be higher in spiral galaxies. However, calculating dynamical equilibrium pressures and comparing them to turbulent pressures in Cen\,A (and how they may relate to the local SFR) is out of the scope of this paper.
}

 \begin{figure*}
 \begin{center}
 \includegraphics[width=0.6\textwidth]{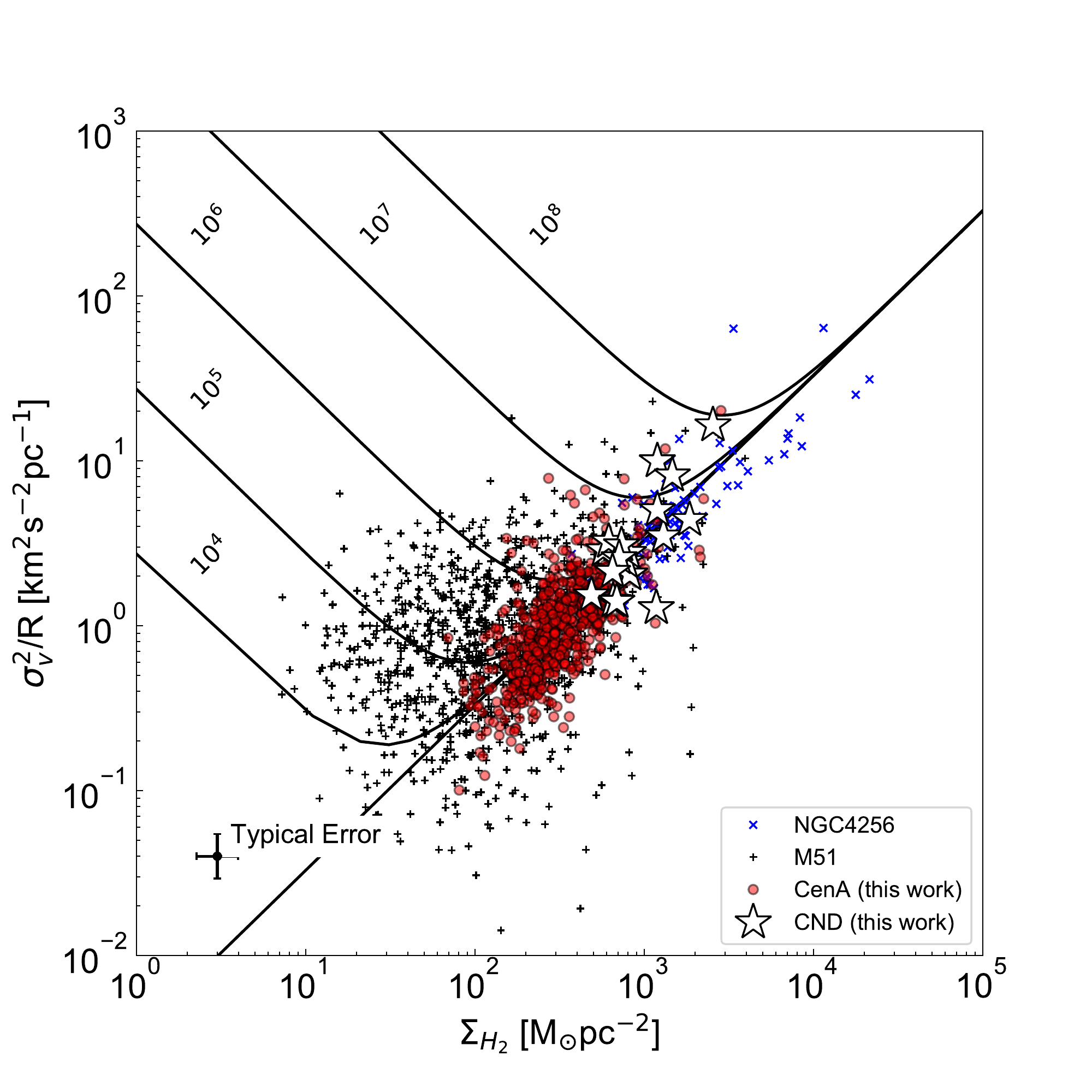}
 \includegraphics[width=0.6\textwidth]{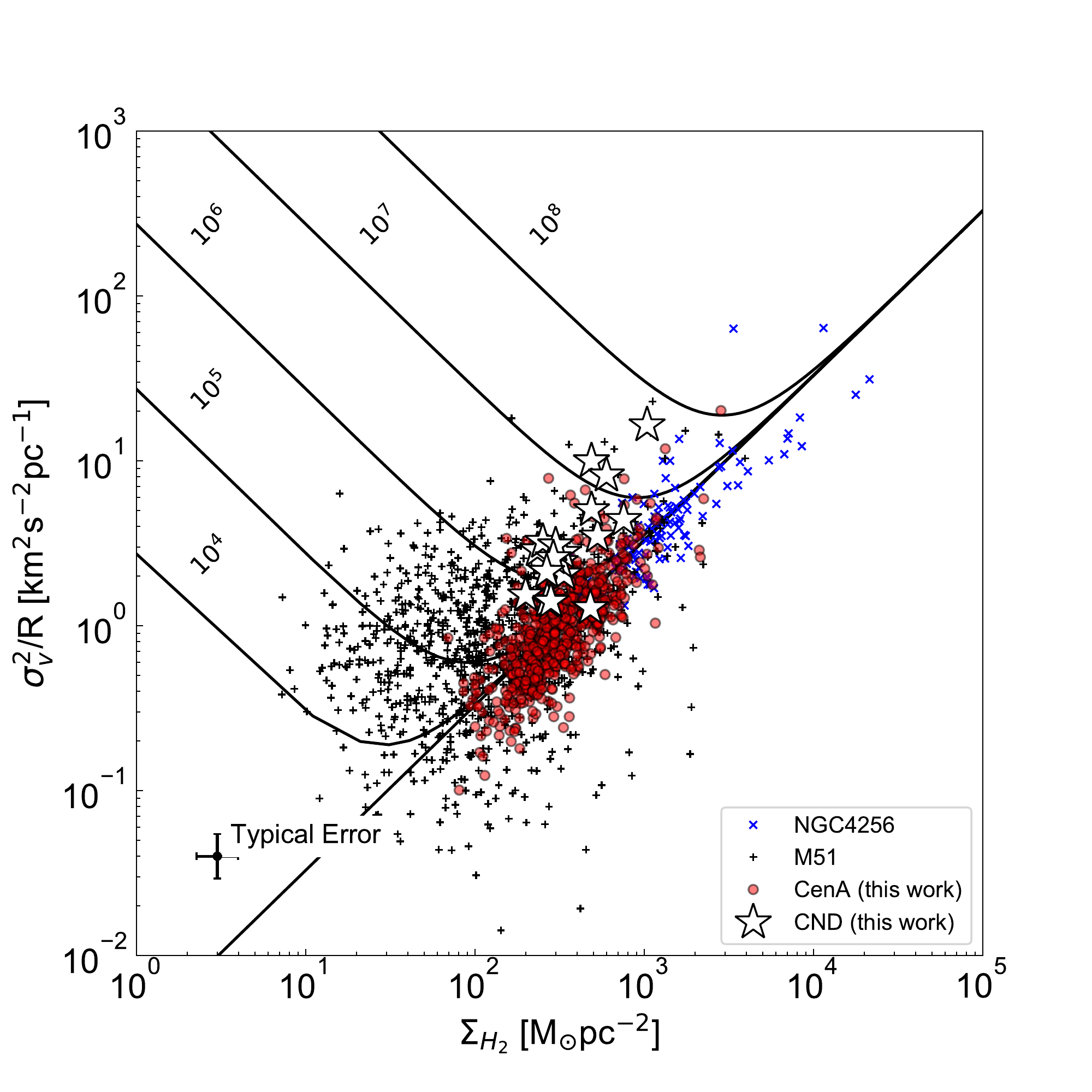}
 \end{center}
 \caption{The scaling coefficient $\sigma_V^2/R$ as a function of surface mass density $\Sigma_{\rm  H_2}$ for the identified GMCs in Centaurus~A.  The white stars represent the GMCs in the CND.  {Top)} For the GMCs in the CND we used $X_{\rm CO}$ = $5\times10^{20}$\,cm$^{-2}$(K\,\kms)$^{-1}$ and for the other regions of the molecular disk  $X_{\rm CO}$ = $2\times10^{20}$\,cm$^{-2}$(K\,\kms)$^{-1}$.
 The solid lines represent equilibrium for external pressures with $P/k_B=0$ (straight line), $10^4, 10^5,10^6, 10^7$, and $10^8$\,cm$^{-3}$\,K \citep{2011MNRAS.416..710F}.
Other symbols are as in Fig.\,\ref{larson}. 
{Bottom) Same as the previous panel, but for the assumption of a constant $X_{\rm CO}$ = $2\times10^{20}$\,cm$^{-2}$(K\,\kms)$^{-1}$ everywhere.}
 \label{pressure}}
 \end{figure*}

 \subsection{ Comparison of Virial Parameter with those in other Early-type Galaxies}

 {In this section we compare the virial parameters ($\alpha_{\rm vir}$ = $M_{\rm vir}$/$M_{\rm gas}$) obtained in \S\,\ref{xcofactor} (see also Fig.\,\ref{virial}) with those obtained for molecular clouds/associations in other early type galaxies, using a common Milky Way disk $X_{\rm CO}$ factor reference.}

 \citet{2015ApJ...803...16U} found that, in the lenticular NGC\,4526, $\alpha_{\rm vir}$ $\simeq$ 1.26 and the standard deviation is $\sim$0.15\,dex, but some clouds close to the galactic center were also characterized by larger values, $\alpha_{\rm vir}$ $\simeq$ 3.5.
 This is in contrast with \citet{2018ApJ...858...17T}, who claimed that in some resolved GMCs (or giant molecular associations) in two elliptical galaxies, NGC\,5846 and NGC\,5044, the GMCs had larger virial parameters ($\alpha_{\rm vir}$$>$14).

  Therefore the slightly higher values in the CND in the case of a common $X_{\rm CO}$ factor would mean unbound conditions probably due to shear in the central regions or caused by other dynamical effects, similar to the case of NGC\,4526 \citep{2015ApJ...803...16U}. At any rate we can discard extreme cases such as those reported by \citet{2018ApJ...858...17T} in the molecular disk of Cen\,A.

 \subsection{ GMC Mass Spectra Across the Molecular Disk of Centaurus A}
 \label{massspectrum}

 \begin{figure*}
 \begin{center}
  \includegraphics[width=0.7\textwidth]{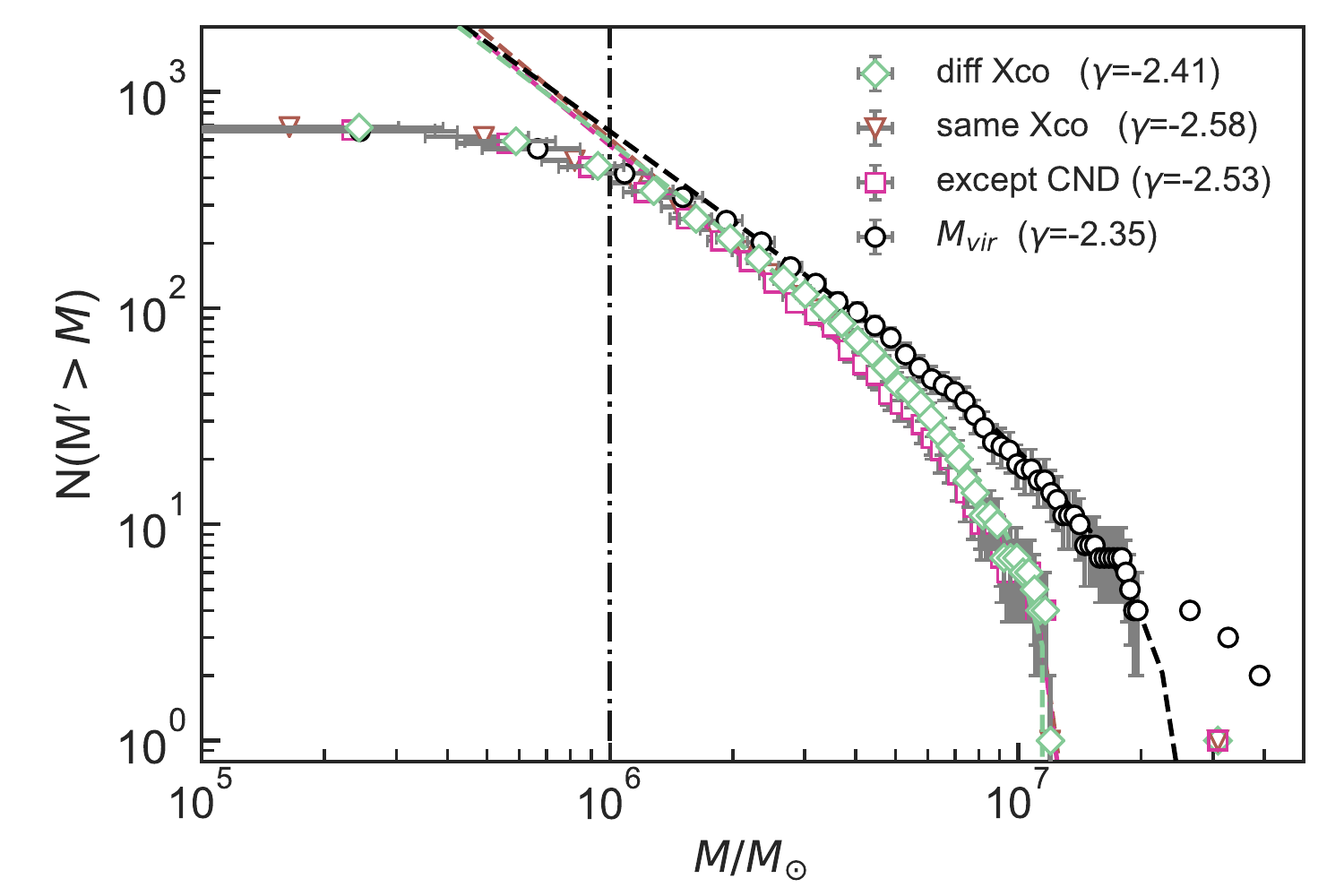}
 \includegraphics[width=0.7\textwidth]{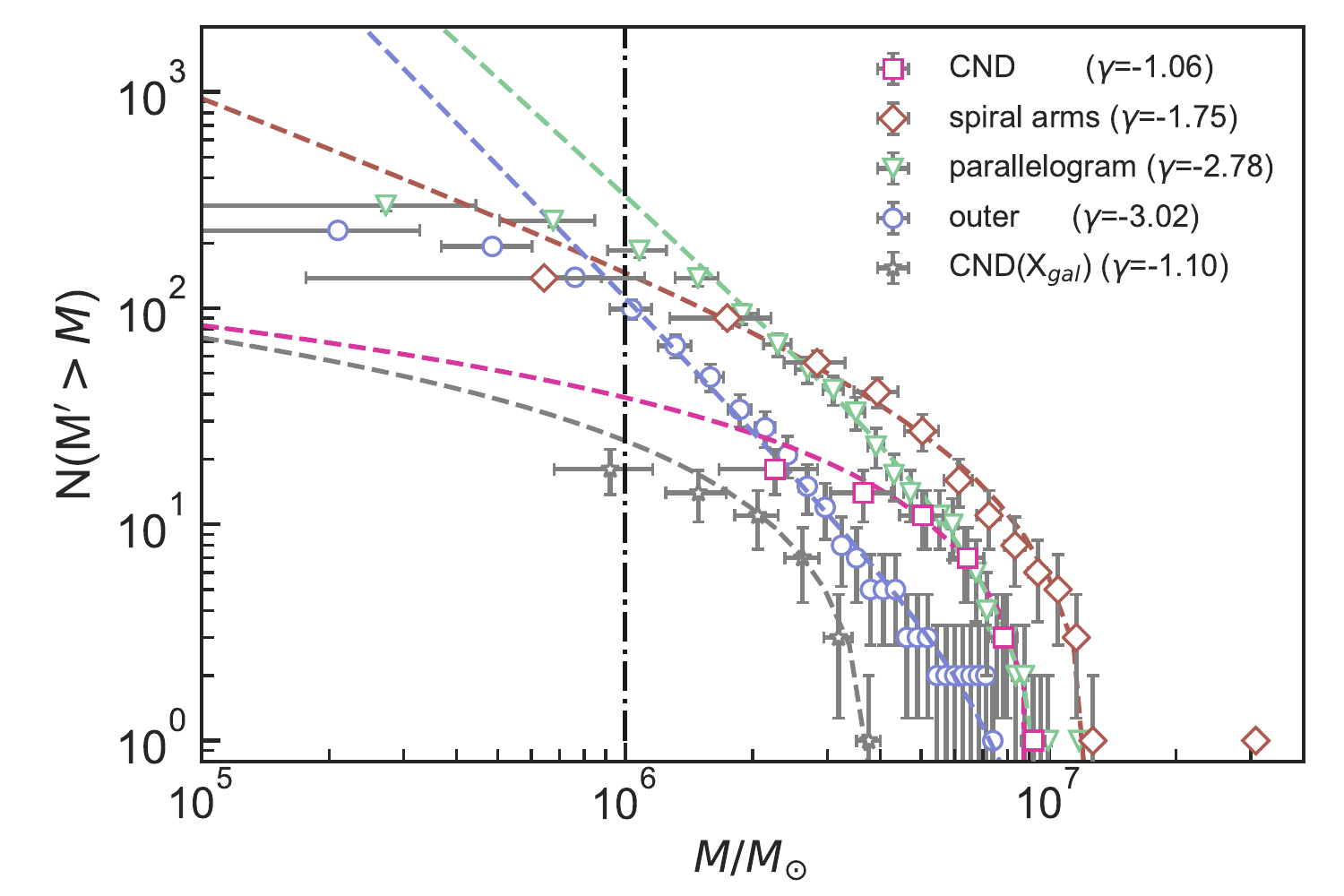}
 \end{center}
 \caption{
 { (Top)} The mass spectra of the GMCs in the molecular disk of Cen~A, obtained:  i) using a conversion factor for the CND region of $X_{\rm CO}$ = $5\times10^{20}$\,cm$^{-2}$(K\,\kms)$^{-1}$ and $X_{\rm CO}$ = $2\times10^{20}$\,cm$^{-2}$(K\,\kms)$^{-1}$ for the other clouds (green), ii) using $X_{\rm CO}$ = $2\times10^{20}$\,cm$^{-2}$(K\,\kms)$^{-1}$ for all GMCs (brown), and iii) excluding the GMCs in the CND (magenta) (see \S\,\ref{massspectrum} for details). The mass spectrum for the derived virial masses is presented as well (black). The vertical line indicates the lower limit of the mass that is used for the fitting, $10^6$\,\Msol. Note that the mass spectra of cases ({\it i})--({\it iii}) are very close to each other and data points/curves overlap.
 { (Bottom)}  The mass spectra of the GMCs belonging to the different regions:  i) CND (magenta, gray), ii) spiral arms (brown), iii) parallelogram structure (green), and iv) outer disk (blue). For the CND we used either $X_{\rm CO}$ = $5\times10^{20}$\,cm$^{-2}$(K\,\kms)$^{-1}$ (magenta) or $X_{\rm CO}$ = $2\times10^{20}$\,cm$^{-2}$(K\,\kms)$^{-1}$ (gray). The color code is the same as in the GMC identification plot in Fig.\,\ref{astrodendro}. 
 \label{massspec}}
 \end{figure*}

 The shape of the GMC mass spectrum is known to vary across the different regions of galaxy disks \citep[e.g.][]{1997ApJ...476..166W,2007ApJ...661..830R}. While in some regions smaller GMCs are predominant due to the destruction of larger clouds caused by stellar feedback and dynamical effects such as shear motions, in other regions a larger population of GMCs may exist due to mechanisms that bring small clouds together such as in the density waves of spiral arms \citep[e.g.][]{1990ApJ...363..435W,2008MNRAS.391..844D,2015ApJ...806...72M,2018PASJ...70...73H}.

 In Fig.\,\ref{massspec} we show the GMC mass spectrum of the molecular disk of Cen~A.
 To determine the optimal bin size, we used an automated bin size selection as implemented in {\sf numpy}.
 The algorithm chooses the `Sturges' estimator \citep{doi:10.1080/01621459.1926.10502161} because it is designed for relatively small data sets ($<1000$ data points). This estimator assumes that the data is distributed as a normal Gaussian distribution and the bin size is defined as $\log_2({n}) +1$, where $n$ is the number of data points.

{There is evidence for significant truncation in the mass distributions at their upper ends (see Fig.\,\ref{massspec} and Table\,\ref{tbl3}).}
 We fitted the cumulative mass function to a truncated power-law form using a completeness mass limit of $10^6$\,\Msol, which is also equivalent to the median mass we obtain for all the identified GMCs in Cen~A.
 The truncated mass function is given by:
 \begin{equation}
 N[M^{\prime}>M] =N_{\rm u}\left[\left(\frac{M}{M_{\rm u}}\right)^{\gamma+1}-1\right],
 \end{equation}
 where $M_{\rm u}$ is the upper cutoff mass, $N_{\rm u}$ the number of GMCs more massive than $2^{1/(\gamma+1)}M_{\rm u}$, and $\gamma$ is the power-law index
 \citep{1997ApJ...476..166W, 2007ApJ...661..830R}.

  We used the bootstrapping method to calculate the uncertainties of the power-law fitting parameters. 
 {First we changed each of the individual GMC masses assuming a normal probability function with a mean $\log M$ and a standard deviation of $0.434 (\delta M/M)$ (where $M$ is the GMC mass and $\delta M$ its uncertainty) and then generated a mass spectrum using the same procedure as explained above.}
 Then we fitted the mass spectrum using the orthogonal distance regression method, taking into account the uncertainties in both the x and y-axis.
 For the x-axis error we use the size of each mass bin, and for the y-axis, the error in each bin is the square root of the number of clouds in that bin.
 We repeated this 10,000 times and used the standard deviation of the fitting parameters of all those simulated histograms as their errors.

 The obtained best-fit parameters are given in Table\,\ref{tbl3}. To calculate the mass spectrum we have considered four different cases: ({\it i}) using different conversion factors to derive the molecular gas masses, i.e. $X_{\rm CO}=5\times10^{20}$\,cm$^{-2}$(K\,\kms)$^{-1}$ for the GMCs in the CND and $X_{\rm CO}=2\times10^{20}$\,cm$^{-2}$(K\,\kms)$^{-1}$ for the rest, ({\it ii}) using the same conversion factor for all GMCs, equal to $X_{\rm CO}=2\times10^{20}$\,cm$^{-2}$(K\,\kms)$^{-1}$, ({\it iii}) including all GMCs except the GMCs in the CND, and ({\it iv}) the virial masses.
  To fit the mass spectrum, {we note that we excluded GMCs with masses of $>1.2\times10^7$\,\Msol\  because the low number statistics to calculate the mean of the high mass end bins (only one GMC per bin) would bias the fit.}
 We find similar index parameters in these four cases, $\gamma\simeq -2.4$ to $-2.6$, which means that $\gamma$ does not strongly depend on the used conversion factor nor usage of CO(1--0) luminosity mass or virial mass.

 We obtained mass spectra for the four distinct regions of the molecular disk as described in \S\,\ref{intro} (see also Paper {\sc i}): ({\it i}) CND, ({\it ii}) spiral arms, ({\it iii}) 'parallelogram' region (i.e. high surface density region as seen in projection), and ({\it iv}) the outermost region of the molecular disk. The GMCs associated with each of these regions are indicated with different color codes in Fig.\,\ref{astrodendro}, and the association of each individual GMC is provided in Table\,\ref{tbl2}. {For the mass distribution fitting, similarly as indicated earlier, we also note that in the arms region we excluded the most massive GMC, with a mass of $3\times10^7$\,\Msol , and in the parallegram region we excluded the second most massive GMC, with a mass of $1.2\times10^7$\,\Msol\ (i.e. just two most massive GMCs are excluded among all).}

There is a trend of steeper mass spectrum shapes with larger radii. The mass spectra of the outermost regions of the molecular disk and the parallelogram structure have a relatively steeper shape ($\gamma = -3.02 \pm 0.08 $ and $ -2.78 \pm 0.03  $, respectively) than that of the spiral arm region ($\gamma=-1.75 \pm 0.05$).

{The obtained parameters and the observed trend are robust. To check how sensitive the parameters of the mass distribution fits are to the completeness/confusion limit, we also obtained these parameters using 0.5 $\times$ 10$^6$\,\Msol\  and 2 $\times$ 10$^6$\,\Msol\ as limits. While the values of the parameters change slightly, these are not substantially large. Using these two limits, the slope changes only by 
$\pm$ 0.1--0.2 for the various assumptions of different Xco, same Xco, all regions except CND, and virial masses.
The trend of the parameters observed across the different regions (from low to high values: outer, parallelogram, spiral arms, CND) also remains largely unchanged for this range of completeness limits.
}

 { We note that in a crowded region, as mentioned in \S\,\ref{methods}, an identified GMC may possibly be blended and composed of multiple smaller GMCs. As a test, we investigate how the mass spectrum would look like if  GMCs were in fact composed of two smaller and equally massive GMCs (e.g a GMC with a mass of $10^7$\,\Msol\ is divided into two GMCs with masses of $5\times10^6$\,\Msol) and then calculate the spectrum index in the same manner.
 In this test we obtained mass spectrum indexes of $-2.23\pm0.03, -2.87\pm0.03, -3.19\pm0.03$, and $-1.11\pm0.20$ for the spiral arms, parallelogram, outer, and CND regions, respectively.
 Although the index tends to be smaller (i.e. fit is steeper) in this test compared to our results, we confirm that the derived spectrum indexes remain similar, and the trend of steeper mass spectrum shapes with larger radii still remains.
 }
 
 {We also note that there must be a break in the mass spectrum of the parallelogram and outer regions whose $\gamma<-2$, because the total CO luminosity estimated from the single index exceeds the actual observed luminosity by more than a factor of two. The break point will likely occur around or below the completeness limit and the steepness would not change by more than +0.5.}

  A steep shape of the mass spectrum indicates that the population of lower mass GMCs is more dominant than that of the most massive clouds.
 The massive GMCs in the parallelogram region and outer disk may not be formed due to a lack of a mechanism that facilitates the agglomeration of molecular clouds. Alternatively massive GMCs may be destroyed by strong stellar radiation fields from young massive stars formed in the molecular disk or from the radiation field of the elliptical galaxy itself. { In fact, the star formation efficiency is seen to be higher toward the outer regions (Paper {\sc I}).
   In the case of the spiral arm regions of Cen~A (shallower mass spectrum of $\gamma=-1.75 \pm 0.05$), massive GMCs are likely formed by collisional agglomeration of smaller clouds in the spiral density wave \citep{2008MNRAS.391..844D}.

 The mass spectrum of the CND is characterized by an even shallower slope, $\gamma = -1.1 \pm 0.2$.
 This may be partly due to the lack of GMCs in the low-mass end range.
  { The total mass of the identified GMCs in the CND region accounts for most (72\,\%) of the total CO luminosity (\S\,\ref{mainGMCprop} and table\,\ref{tbl4}), so the  diffuse and extended molecular component is not likely to be a dominant contributor. Even if the low-mass end of GMCs is not completely traced, the number of GMCs with masses $\sim10^6$\,\Msol\ would not exceed 40, which is close to the prediction from the fit.
 {The impact of missing the lower mass GMCs can also be examined by integrating the fitted mass spectra \citep{2018PASJ...70...73H}.
 The ratio of cloud mass integrated from a certain low mass limit $M_{\rm low}$ to the highest cloud mass $M_{\rm high}$, and the total mass from $M_{\rm low}$ = 0 is given by $(1 - (M_{\rm low} / M_{\rm high})^{\gamma + 2}$)  \citep[see Eq.\,15 in][]{2018PASJ...70...73H}.
 The lowest and highest mass limits for the CND mass spectra are $2\times10^6$\,\Msol\ and $10^7$\,\Msol, respectively (see Fig.\,\ref{massspec}). Then, for $\gamma=-1.1$ the ratio is $\sim0.76$, which is close to 72\,\%. }
  Thus the above mentioned 72\% of molecular gas in GMC form indicate that the spectral index would not significantly decrease due to non-detections of low-mass GMCs.
 We also note that we had excluded 5 GMCs with $\sim3\times10^6$\,\Msol\ in the central region within a velocity range between 534 and 564\,\kms\ (\S\,\ref{methods}). However, even adding these clouds, the low-mass end would still be significantly deficient compared to the other regions. }

 The shallower mass spectrum of the CND may be partly related to the agglomeration of molecular clouds along the observed molecular filaments, but the situation is more complex because other mechanisms are likely playing a role as well. The massive GMCs that originally formed at the arm regions and migrated to the CND (gas collides and loses angular momentum - unlike the stars) might be disrupted. Strong shocks with large shear motions may be an important mechanism to destroy the largest clouds in the CND although we note that these shocks are located within the inner 100 pc \citep{2017ApJ...843..136E}. Also, small ($<$10$^5$\,$M_\odot$) molecular clumps and unbound diffuse molecular gas may have been selectively destroyed due to strong radiation by the AGN, while massive and denser clouds can resist longer such effect \citep{2001A&A...377.1016V,2010A&A...522A..24H,2011A&A...536A..41H,2014MNRAS.443.2018N}.
 }

 \subsubsection{Comparison with GMC mass spectra in other galaxies}
 Finding different shapes of the mass spectra across different regions in a given galaxy is not uncommon in the literature.
 Note that here we only compare our results with studies where maps have a similar spatial resolution ($\sim$ a few 10\,pc) and the same cloud identification method. This is because the shapes of the mass spectra can be largely biased by the method used to decompose the clouds \citep{2015MNRAS.454.2067C}.

 Steeper shapes of the mass spectrum ($\gamma < -2 $) are found in another early-type galaxy (although of lenticular S0 type), NGC\,4526 \citep[$\gamma=-2.39$;][]{2015ApJ...803...16U}, in the outer disk of spiral galaxies \citep[$\gamma	\lesssim -2.3$ to --2.6;][see also \citealt{2016ApJ...822...52R}]{2005PASP..117.1403R,2012A&A...542A.108G} and in the inter-arm regions \citep[$\gamma\sim -2.5$;][]{2014ApJ...784....3C},  as well as in the Large Magellanic Cloud \citep[$\gamma<-2$;][]{2011ApJS..197...16W}.
  On the other hand, shallower slopes are found in the spiral arms of M\,51 \citep[$\gamma\sim-1.8$;][]{2014ApJ...784....3C}, M\,33 and our Galaxy \citep[$\gamma\sim-1.4\,$ to $ \,-1.6$;][]{2012A&A...542A.108G,2005PASP..117.1403R}. Therefore the newly formed spiral arms of Cen~A already have similar properties as those of late type spiral galaxies.

 In the case of Cen~A the cutoff mass is about $10^7$\,\Msol\ at intermediate radii although we note that the most massive GMCs ($>10^7$\,\Msol) are found toward the molecular spiral arm region.
 This cutoff mass is similar  to that of the GMCs in the disk regions of spiral galaxies such as M\,51 \citep{2014ApJ...784....3C}. However, it doubles that of the S0 galaxy NGC\,4526,  $M_{\rm u}$ = 4 $\times$ 10$^6$\,\Msol\ \citep{2015ApJ...803...16U}.
 {The most massive GMC is in the spiral arm feature, and has a size of around 91\,pc, width of 20\,km/s, exceeding by a factor of two or three the median of all GMCs in the molecular disk of Cen\,A, and stands out by a factor of more than 25 in CO(1-0) luminosity. It has similar characteristics to the largest GMC found in NGC\,628, which has survived and grown probably because it is located at the intersection of the co-rotation radius and one of its spiral arms \citep{2020A&A...634A.121H}}.
 This confirms that in Cen~A the spiral pattern is an important mechanism where smaller clouds are aggregated to form larger GMCs, similarly to late type disk galaxies, and this mechanism might not be present in NGC\,4526, or other destructive mechanisms are more dominant.
 In fact, \citet{2015ApJ...803...16U} argue that in NGC\,4526 the properties of GMCs might be dominated by a high stellar radiation field (although this situation is probably also common to Cen~A), which may destroy large molecular clouds. 
 It is probably in the less massive and lower surface density GMCs located at the disk outskirts of Cen~A where the mass spectrum shape is most similar to that in NGC\,4526, probably due to GMCs being similarly affected by that mechanism.

  We found that the mass spectrum index of the CND in Cen~A is $\gamma = -1.1 \pm 0.2$.
 A shallow mass spectrum ($\gamma=-1.4$) and smaller cutoff mass (1.9 $\times$ 10$^6$ $M_\odot$) were found in the inner regions of NGC\,4526 compared to other regions \citep[][]{2015ApJ...803...16U}.
 The situation is similar in the nuclear bar of M51 where the mass spectrum presents an index of  $\gamma = -1.3$, and a similar truncation for cloud masses above $M$ $\sim$ 5.5 $\times$ 10$^6$ $M_\odot$ \citep{2014ApJ...784....3C}. In both studies a constant $X_{\rm CO}$ = 2$\times10^{20}$\,cm$^{-2}$\,{(K\,\kms)$^{-1}$ was assumed.

  With the high linear resolution of a few 10\,pc scale used here, only a few other studies focused on a similar GMC identification and the mass spectrum in regions close to an AGN. In addition, they are mostly low luminosity AGNs and not as long-lived as in Cen~A given the large extent of the radio source. In M51's nuclear bar region a combination of mechanisms might be contributing to cloud disruption and heating of the molecular gas, but it is difficult to separate their individual contribution in the observed mass spectrum \citep{2014ApJ...784....3C}.

 In the case of Cen~A we favor a different $X_{\rm CO}$ factor for the inner and outer regions.
 However, if on the other hand we assume the same $X_{\rm CO}$ factor (i.e. $X_{\rm CO}$ = 2$\times10^{20}$\,cm$^{-2}$\,(K\,\kms)$^{-1}$) for the CND and for other regions in the molecular disk, the truncation in the mass spectrum of the CND is smaller than in the other regions, a similar trend to that seen in NGC\,4526 and M51.
 The maximum mass in the CND would then be $4.0\times10^6$\,\Msol.
 Another consequence of this assumption would be that the GMCs will not be in virial equilibrium and would need to be supported by external pressure.


 \subsubsection{Comparison with GMC mass functions in the context of numerical calculations }

 { Next we compare our observed GMC mass spectra with existing numerical calculations in the literature.
 \citet{2017ApJ...836..175K} proposed a semi-analytical time evolution model of GMC mass functions by considering multiple processes, i.e. cloud formation from a magnetized interstellar medium through multiple episodes of compression by \ion{H}{ii} regions and supernova remnants, cloud dispersal due to stellar feedback by massive stars, cloud-cloud collisions, and gas recycled to regenerate or grow pre-existing GMCs. }
 In the case of relatively low to mid GMC masses $M$ $<$ 10$^{5.5}$\,$M_\odot$ the cloud-cloud collision terms are negligible, and thus the slope of the power-law can be approximated by $dN/dM \propto M^{-1-T_{\rm f}/T_{\rm d}}$, where $T_{\rm d}$ is the typical dispersal time and $T_{\rm f }$ is the typical formation/mass-growth timescale of GMCs.

 Applying this to the aforementioned regions with steeper mass spectra, those outside Cen~A's CND, the molecular cloud formation timescale would be relatively long compared to the destruction timescale in the parallelogram structure and outer disk, while the formation timescale would be relatively short in the spiral arm region.
 The situation of arms, parallelograms, and outer regions can be explained within this framework, where GMC mass spectrum is formed due to the balance between formation process and destruction process. However, the situation in the CND is less clear.
 To be able to reproduce an index for the CND of $\gamma = -1.1$, the dispersal time would have to be 10 times longer than the formation timescale. However due to the radiation and energetics close to an AGN, the dispersal time would be expected to be shorter (i.e. strong and fast impact on cloud disruption) than under normal conditions.
 { The dynamical timescale is short in the CND, and the destruction timescale due to shear is also presumably short \citep[e.g., $\sim 1 - 4$\,Myr;][]{2018MNRAS.478.3380J}.

  Since the number density of GMCs is likely higher in the CND than in the outer regions, cloud-cloud collisions may dominate other processes (e.g., ISM phase transitions driven by \ion{H}{ii} regions and/or supernova remnants). However, such collisional processes easily produce steep slopes because the mass-growth rate due to collisions is larger for more massive clouds \citep{2017ApJ...836..175K}. A shallow slope of $\gamma = -1.1$ cannot be reproduced unless the collisional kernel, which governs the collision rate, has almost no mass-dependence \citep{2018PASJ...70S..59K}.

 All these points indicate that, to explain the observed flat spectrum, destruction in massive clouds has to be less efficient in the CND than in other regions so that $T_{\rm d}$ is somehow longer in massive clouds.
 In other words, in the CND the conditions must be considerably different to that in the arms and inter-arm regions of disk galaxies and not easily reproducible with numerical calculations currently available in the literature.
 We note that the shallower cloud mass spectrum in the circumnuclear regions is not simply a question related to the AGN because this trend is often found in centers of other disk galaxies \citep[e.g.][]{2014ApJ...784....3C}.

 To understand the mechanism that shapes the mass spectrum in the CND, further models including AGN activity are needed. These should take into account how much gas falls into the CND via the arms, how much is lost or entrained by the action of the jet \citep[e.g.][]{ 2012ApJ...757..136W}, and how much efficiently the GMCs are destroyed due to the impact of radiation and winds \citep[e.g.][]{2013ApJ...763L..18W}.
  }

 \section{Summary}

 We present the first census of Giant Molecular Clouds (GMCs) { complete down to 10$^6$\,$M_\odot$} and within the inner 4\,kpc} of the molecular disk of the nearest giant elliptical and powerful radio galaxy, Centaurus~A. This is obtained by means of
 high angular/spectral resolution and high sensitivity ALMA CO(1--0) data. We combined ALMA 12m, 7m and TP array data in order to have complete information from small to large spatial scales and recover all the flux. We have successfully resolved  the molecular disk of Centaurus~A into tens of parsec scale GMCs using CPROPS.  Our main results are:

 \begin{enumerate}

 \item We have identified a total of 689 GMCs across the dust lane of Centaurus~A. They are characterized by a median size of 38\,pc. The median velocity dispersion is 6.1\,\kms , while in the CND they are characterized by larger velocity width of 12.4\,\kms .

 \item We found that the GMCs in Centaurus~A are offset by 0.14\,dex from the general line width - size relation found in nearby galaxies and the Galactic disk. GMCs in the CND systematically present the largest offsets, 0.43\,dex in average.

 \item We have obtained the $X_{\rm CO}$ factor using the virial method for the first time in this object. It is $X_{\rm CO}= (2 \pm 1)\times10^{20}$\,cm$^{-2}$(K\,\kms)$^{-1}$ in the molecular disk. In the CND we find instead $X_{\rm CO}$ = $(5 \pm 2 ) \times10^{20}$\,cm$^{-2}$(K\,\kms)$^{-1}$. The larger value of the CND is in good agreement with a previous independent measurement by \citet{2014A&A...562A..96I}. It is not likely that this is due to a metallicity dependence because the metallicity has been shown to be almost constant across the molecular disk \citep[0.7--0.8\,$Z_\odot$; ][]{2017A&A...599A..53I}.

 \item GMCs are located along a line of surface density of $\Sigma_{\rm H_2}$ $\sim$ 300\,$M_\odot$\,pc$^{-2}$, higher than the general trend for the molecular clouds in our Galaxy and other nearby galaxies, but similar to those in the Galactic Center. In general, external pressure is not needed to support the GMCs (if $X_{\rm CO} = 5 \times10^{20}$\,cm$^{-2}$\,(K\,\kms)$^{-1}$ near the center) and can be gravitationally bound.

 \item We obtained the GMC mass spectrum and found that the best fit of a truncated power law for the entire molecular disk is consistent with that found in other disk (spiral and lenticular) galaxies ($\gamma \simeq -2.41 \pm 0.02$, upper cutoff mass $M_{\rm u}$ $\sim 1.3\times10^{7}$\,\Msol). However, in the arms and the CND the fitted curves are shallower, with indices of $\gamma = -1.75 \pm 0.05$ and $-1.1 \pm 0.2$, respectively.

 \item { The different shapes of the mass spectra in the outer regions of the molecular disk and CND of Cen~A, transit from steep to shallow as we move from outer to inner radii. This implies that the properties of GMCs are transformed when they flow from the outer to the central regions.
 In the arms massive GMCs are likely formed by collisional agglomeration of smaller clouds in the spiral density wave.  In the CND the massive GMCs that originally formed in the arm regions and migrated to the CND might be disrupted by the effect of the AGN and  intense shear. Other competing mechanisms such as AGN radiation can be disrupting lower mass GMCs, resulting in the shallow shape of the mass spectrum.}

 \end{enumerate}


              \begin{table*}
              \begin{center} 
             
              \caption{Properties of Giant Molecular Clouds in the Molecular Disk of Centaurus\,A \label{tbl2}}
              \begin{tabular}{ccccccrrrrrrrcr}
              \hline
              ID &($\Delta\alpha, \Delta\delta$)\textsuperscript{a} & $v_{\rm LSR}$ & $\sigma_v$ & 
              $\sigma_{\rm maj}\times \sigma_{\rm min}$ (P.A.)\textsuperscript{b} & $R$\textsuperscript{c} & 
              $S_{\rm CO(1-0)}$& $M_{\rm vir}$\textsuperscript{c} &  Region\textsuperscript{d} \\
              & ($\arcsec$, $\arcsec$) &  (\kms) & (\kms) & (pc) & (pc)  & (Jy\,\kms)& (10$^4\,M_{\sun}$)
               \\
\hline
1 & ($97.2, -47.4$) & 259 & 5.3 $\pm$ 0.7 & $23\times19$ ($33$\arcdeg) & 36 $\pm$ 3 &  8.8 $\pm$  1.0 & 107 $\pm$ 28 & P\\ 
2 & ($99.0, -44.3$) & 261 & 4.8 $\pm$ 0.5 & $29\times18$ ($-70$\arcdeg) & 40 $\pm$ 4 &  9.9 $\pm$  0.8 & 99 $\pm$ 24 & P\\ 
3 & ($103.9, -51.5$) & 266 & 6.4 $\pm$ 1.3 & $22\times17$ ($5$\arcdeg) & 32 $\pm$ 5 &  6.0 $\pm$  1.3 & 136 $\pm$ 68 & P\\ 
4 & ($89.0, -30.9$) & 270 & 5.0 $\pm$ 0.7 & $29\times13$ ($87$\arcdeg) & 32 $\pm$ 3 &  7.3 $\pm$  0.9 & 86 $\pm$ 26 & P\\ 
5 & ($101.4, -48.0$) & 266 & 6.9 $\pm$ 0.5 & $36\times19$ ($23$\arcdeg) & 46 $\pm$ 3 & 21.9 $\pm$  1.4 & 232 $\pm$ 42 & P\\ 
6 & ($106.3, -58.1$) & 275 & 7.1 $\pm$ 0.6 & $56\times46$ ($-16$\arcdeg) & 95 $\pm$ 7 & 29.1 $\pm$  1.7 & 505 $\pm$ 107 & P\\ 
7 & ($87.3, -26.3$) & 274 & 9.5 $\pm$ 0.5 & $39\times29$ ($3$\arcdeg) & 62 $\pm$ 2 & 50.5 $\pm$  1.9 & 581 $\pm$ 68 & P\\ 
8 & ($98.3, -47.2$) & 276 & 7.3 $\pm$ 0.6 & $28\times22$ ($3$\arcdeg) & 44 $\pm$ 3 & 13.0 $\pm$  0.9 & 249 $\pm$ 52 & P\\ 
9 & ($90.4, -34.3$) & 273 & 4.8 $\pm$ 0.9 & $23\times15$ ($73$\arcdeg) & 31 $\pm$ 4 &  6.5 $\pm$  0.9 & 77 $\pm$ 30 & P\\ 
10 & ($112.6, -57.7$) & 278 & 6.2 $\pm$ 1.1 & $37\times17$ ($-9$\arcdeg) & 44 $\pm$ 7 & 10.4 $\pm$  1.4 & 176 $\pm$ 65 & P\\ 
...\\
\hline

\multicolumn{9}{p{14cm}}{For details about how the parameters were calculated please refer to \S\,\ref{methods}. The GMC IDs are in order of increasing velocity (in the Local Standard Rest frame). Table\,\ref{tbl2} is published in its entirety online in machine-readable format. The first ten lines are shown here for guidance regarding the format and content.}\\
\multicolumn{9}{p{14cm}}{\textsuperscript{a} \footnotesize{Intensity-weighted peak position relative to the AGN position at $\alpha_=13^{\rm h}25^{\rm m}27\fs615s$, $\delta=-43$\arcdeg01\arcmin08\farcs805.}}\\
\multicolumn{9}{p{14cm}}{\textsuperscript{b} \footnotesize{Major and minor axes of the GMCs without beam deconvolution. The position angles are indicated inside the parentheses, measured counterclockwise from north to east.}}\\
\multicolumn{9}{p{14cm}}{\textsuperscript{c} \footnotesize{Radius and virial masses are not presented for the GMCs whose minor axis is too small to calculate a deconvolved minor axis.}}\\
\multicolumn{9}{p{14cm}}{\textsuperscript{d} \footnotesize{Identification code of the region where the GMC is located (C: CND, S: Spiral arms, P: Parallelogram, O: Outer disk) The C${\ast}$ stands for the five excluded GMCs in the CND that fall within a velocity range between 534 and 564 \kms).}}
\end{tabular}
\end{center} 
\end{table*}

\begin{table*}
\centering
\begin{center} 
\caption{Total CO luminosities compared to those in GMCs\label{tbl4}}
              \begin{tabular}{crrrr}

\hline 
{} & {Number of GMCs} & {Total CO luminosity}  & \multicolumn{2}{c}{CO luminosity in GMCs}   \\ 
{} & {}                             &  {($10^6$\,K\,\kms\,pc$^2$)}&   \multicolumn{2}{c}{($10^6$\,K\,\kms\,pc$^2$)}               \\
\hline
{} & {}                             & {} &{Extrapolation\textsuperscript{a}}  & {No extrapolation\textsuperscript{b}}                                      
\\
\hline 
CND        	    &  23 &  13.4 &        9.6 ( 72\,\%) &    4.4 ( 33\,\%)  \\
Arms 	   & 138 &  71.4 &    89.7 (126\,\%) &   57.5 ( 81\,\%) \\ 
Parallel        & 299 & 134.9 &   108.1 ( 80\,\%) &   58.5 ( 43\,\%) \\ 
Outer    	   & 229 & 127.3 &    55.4 ( 44\,\%) &   25.1 ( 20\,\%)  \\\hline
Total            & 689 & 347.0 &   262.8 ( 76\,\%) &  145.6 ( 42\,\%)      \\                                   
\hline 

\multicolumn{5}{p{10cm}}{\textsuperscript{a} The CO(1--0) luminosity from all the GMCs in each region with flux extrapolation down to zero-intensity (see \S\,\ref{methods} and \S\,\ref{mainGMCprop}).}\\
\multicolumn{5}{p{10cm}}{\textsuperscript{b} The CO(1--0) luminosity from all the GMCs in each region without flux extrapolation down to zero-intensity (above 2\,$\sigma$ level.)}\\

\end{tabular}
\end{center} 
\end{table*}

\clearpage
\begin{table*}{} 
\caption{Parameters of the Power-Law  Mass Distribution Function Fits \label{tbl3}} 
              \begin{tabular}{p{3cm}cccc}

\hline
{~} & {$\gamma$} & {$N_{u}$} & {$M_{u}$}  \\ 
   &          &         & ($10^6$\,\Msol) \\
\hline 
Different Xco     & $ -2.41 \pm  0.02 $ & $  16.7 \pm  0.8 $ & $ 12.8 \pm  0.1 $ \\ 
Same Xco         & $ -2.58 \pm  0.02 $ & $  10.3 \pm  0.6 $ & $ 13.5 \pm  0.2 $ \\ 
Except CND      & $ -2.53 \pm  0.02 $ & $  10.9 \pm  0.6 $ & $ 13.4 \pm  0.2 $ \\ 
$M_{vir}$          & $ -2.35 \pm  0.04 $ & $    7.9 \pm  1.4 $ & $  26.8 \pm  2.1 $ \\  \hline
Arms     & $ -1.75 \pm  0.05 $ & $ 24.6 \pm 3.3$ & $   13.2 \pm  0.2 $ \\ 
Parallel & $ -2.78 \pm  0.03 $ & $  5.3 \pm  0.4 $ & $  10.3 \pm  0.2 $ \\ 
Outer    & $ -3.02 \pm  0.08 $ & $  1.0 \pm  0.3 $ & $ 10.2 \pm  0.8 $ \\ \hline
CND & \textsuperscript{a}$ -1.06 \pm  0.24 $ & $ 263 \pm 106 $ & $  9.8 \pm  0.6 $ \\ 
 ~      & \textsuperscript{b}$ -1.10 \pm  0.27 $ & $ 164 \pm 68 $ & $  4.0 \pm  0.2 $\\
\hline
 
\multicolumn{4}{p{9cm}}{\textsuperscript{a} An $X_{\rm CO}$ factor of $5\times10^{20}$\,cm$^{-2}$(K\,\kms)$^{-1}$ is used for the calculation of the GMC masses.}\\
\multicolumn{4}{p{9cm}}{\textsuperscript{b} With an $X_{\rm CO}$ factor of $2\times10^{20}$\,cm$^{-2}$(K\,\kms)$^{-1}$.}\\

\end{tabular}
\end{table*}

\section*{Acknowledgements}
{We sincerely thank the referee (Erik Rosolowsky) for the careful reading and useful comments to improve our manuscript. }
We would also like to show our gratitude to him for the kind assistance with the usage of CPROPS in the early stages of this work.
R.E.M was supported by the ALMA Japan Research Grant of NAOJ ALMA Project, NAOJ-ALMA-222.
D.E. was supported by JSPS KAKENHI Grant Number JP 17K14254.
D.E. was supported by the ALMA Japan Research Grant of NAOJ ALMA Project, NAOJ-ALMA-0093.
M.I.N.K. was supported by JSPS KAKENHI Grant Number JP 15J04974. K.K. was supported by JSPS KAKENHI Grant Number JP17H06130 and the NAOJ ALMA Scientific Research Grant Number 2017-06B.
S.V. acknowledges support by the research projects AYA2014-53506-P and
AYA2017-84897-P from the Spanish Ministerio de Economia y
Competitividad, from the European Regional Development Funds (FEDER) and
the Junta de Andalucia (Spain) grants FQM108. This study has been
partially financed by the Consejer\'{i}a de Conocimiento, Investigaci\'{o}n y
Universidad, Junta de Andaluc\'{i}a and European Regional Development Fund
(ERDF), ref. SOMM17/6105/UGR.
Part of this work was achieved using the grant of Visiting Scholars
Program supported by the Research Coordination Committee, National
Astronomical Observatory of Japan (NAOJ), National Institutes of Natural
Sciences (NINS). S.M. would like to thank the Ministry of Science and Technology (MOST)
of Taiwan, MOST 107-2119-M-001-020.

This research has made use of NASA's Astrophysics Data System.
This research has made use of Astropy, a community-developed core Python (http://www.python.org) package for Astronomy \citep{2013A&A...558A..33A, 2018AJ....156..123A}; ipython \citep{PER-GRA:2007}; matplotlib \citep{Hunter:2007}; APLpy, an open-source plotting package for Python \citep{2012ascl.soft08017R} and NumPy \citep{5725236}.
Data analysis was in part carried out on the open use data analysis computer system at the Astronomy Data Center, ADC, of the National Astronomical Observatory of Japan.
This research has made use of the NASA/ IPAC Infrared Science Archive, which is operated by the Jet Propulsion Laboratory, California Institute of Technology, under contract with the National Aeronautics and Space Administration.

This paper makes use of the following ALMA data:
ADS/JAO.ALMA\#2013.1.00803.S.
ALMA is a partnership of ESO (representing its member states), NSF (USA) and NINS (Japan), together with NRC (Canada), MOST and ASIAA (Taiwan), and KASI (Republic of Korea), in cooperation with the Republic of Chile. The Joint ALMA Observatory is operated by ESO, AUI/NRAO and NAOJ.
The National Radio Astronomy Observatory is a facility of the National Science Foundation operated under cooperative agreement by Associated Universities, Inc.

\section*{Data Availability}
{This paper makes use of the following ALMA data:
ADS/JAO.ALMA\#2013.1.00803.S. 
All original data is accessible from the ALMA Science Archive. 
The processed data underlying this article will be shared on reasonable request to the corresponding author.
}



\bibliographystyle{mnras}
\bibliography{cenaGMCs}




\appendix




\bsp	
\label{lastpage}
\end{document}